\documentclass[12pt,a4paper]{article}
\usepackage{amsmath,amssymb,mathrsfs,framed,esint,slashed}
\usepackage[colorlinks]{hyperref}
\usepackage{color}
\usepackage[all]{xy}

\newtheorem{theorem}{Theorem}[section]
\newtheorem{definition}[theorem]{Definition}
\newtheorem{lemma}[theorem]{Lemma}

\newtheorem{proposition}[theorem]{Proposition}
\newtheorem{corollary}[theorem]{Corollary}

\newcommand{\rem}[1]{}

\newcommand{\de}{{\rm d}}

\newcommand{\R}{{\mathcal{R}}}

\newcommand{\bx}{{\boldsymbol{x}}}

\newcommand{\bJ}{{\mathbf{J}}}

\newcommand{\beq}{\begin{equation}}
\newcommand{\eeq}{\end{equation}}
\newcommand{\bal}{\begin{align}}
\newcommand{\eal}{\end{align}}

\numberwithin{equation}{section}

\newcommand{\todo}[1]{\vspace{5 mm}\par \noindent
\framebox{\begin{minipage}[c]{0.95 \textwidth}
\tt #1 \end{minipage}}\vspace{5 mm}\par}

\begin{document}

\title{\vspace{-.7cm}Madelung transform and probability densities\\in hybrid quantum--classical dynamics}
\author{Fran\c{c}ois Gay-Balmaz$^1$, Cesare Tronci$^{2,3}$ \smallskip 
\\ 
\footnotesize
\it $^1$CNRS and \'Ecole Normale Sup\'erieure, Laboratoire de M\'et\'eorologie Dynamique, Paris, France
\\
\footnotesize
\it $^2$Department of Mathematics, University of Surrey, Guildford, United Kingdom
\\
\footnotesize
\it 
$^3$Numerical Methods Division, Max Planck Institute for Plasma Physics, Garching, Germany}
\date{$\,$}
\maketitle

\vspace{-1.4cm}
\begin{abstract} \footnotesize
This paper extends the Madelung-Bohm formulation of quantum mechanics to describe the time-reversible interaction of classical and quantum systems. The symplectic geometry of the Madelung transform leads to identifying hybrid quantum--classical Lagrangian paths extending the  Bohmian trajectories from standard quantum theory. As the classical symplectic form is no longer preserved, the nontrivial evolution of the Poincar\'e integral is presented explicitly. Nevertheless, the classical phase-space components of the hybrid Bohmian trajectory identify a Hamiltonian flow parameterized by the quantum coordinate and this flow is associated to the motion of the classical subsystem.
In addition, the continuity equation of the joint quantum--classical  density is presented explicitly. While the von Neumann density operator of the quantum subsystem is always positive-definite by construction, the hybrid  density is generally allowed to be unsigned. However, the paper concludes by presenting an infinite family of hybrid Hamiltonians whose corresponding evolution preserves the sign of  the probability density for the classical subsystem.

\end{abstract}

\vspace{-.6cm}
{\scriptsize

\tableofcontents\tiny
}

\section{Introduction}

This paper deals with the dynamics of coupled classical and quantum degrees of freedom. This topic has been attracting much attention since the early speculations on the role of the classical apparatus in the theory of quantum measurement \cite{VonNeumann,WhZu}. In the usual approach, one starts with a full quantum treatment for all degrees of freedom and then takes the semiclassical limit on some of them. Over the decades, this approach has led to several models differing in the way the semiclassical limit is performed. On the other hand, the alternative approach followed in the present work seeks a mathematically consistent description of hybrid quantum--classical systems that are not necessarily the limit of a fully quantum theory. In other words, classical motion is not regarded in this framework as an approximation of quantum mechanics. While this construction has led to the celebrated {\it quantum--classical Liouville equation} \cite{Aleksandrov,boucher,Gerasimenko}  in chemical physics \cite{Kapral}, this equation suffers from the essential drawback of not preserving the quantum uncertainty relations. Indeed, the quantum--classical Liouville equation generally allows the quantum density matrix to become unsigned \cite{Diosi}. Other hybrid theories \cite{PrKi}  also suffer from similar issues.  Within the context of Hamiltonian dynamics, alternative  theories also exist. In some cases \cite{Hall} they retain quantum--classical correlations even in the absence of coupling. 
In some other cases \cite{Sudarshan}, the emergence of further interpretative issues   \cite{Barcelo,PeTe} led some to exclude the possibility of a mathematically and physically consistent theory of quantum--classical coupling \cite{Terno,Salcedo}. Nevertheless, the search for a model of quantum--classical correlation dynamics is currently still open.

Recently, a hybrid quantum--classical wave equation was formulated in \cite{BoGBTr} by using momentum map methods in symplectic geometry \cite{Sternberg2,MaRa} so that the system naturally inherits a standard Hamiltonian structure. General symplectic methods have been continuously successful in quantum theory \cite{deGosson}, while momentum maps in geometric mechanics have been shown to be  particularly advantageous for Gaussian quantum states \cite{BLTr2015,BLTr2016,OhTr2017,OhLe}, quantum hydrodynamics \cite{FoHoTr}, and mixed state dynamics \cite{Montgomery,Tronci2018}.  The Hamiltonian quantum--classical theory in \cite{BoGBTr} was largely inspired by the crucial contribution by George Sudarshan \cite{Marmo,Sudarshan,Sudarshan2}, who in 1976 proposed  to describe hybrid quantum--classical systems by exploiting the Koopman-von Neumann (KvN)  formulation of classical mechanics \cite{Koopman,VonNeumann2}. The KvN theory proposes to describe classical mechanics in terms of wavefunctions, thereby allowing for a common Hilbert-space framework which is then shared by both classical and quantum mechanics.
 In the KvN construction, the classical probability density $\rho(q,p)$ is represented in terms of a wavefunction $\Psi(q,p)$ by setting $\rho=|\Psi|^2$. A direct verification shows that if $\Psi$ satisfies the KvN equation
\begin{equation}
{\rm i}\hbar\partial_t\Psi=\widehat{L}_H\Psi
\,,\qquad\text{with}\qquad
\widehat{L}_H:={\rm i}\hbar\{H,\ \}
,
\label{KvNeq}
\end{equation}
then $\rho=|\Psi|^2$ satisfies the Liouville equation $\partial_t\rho=\{H,\rho\}$ from classical mechanics.
Here, $H$ is the Hamiltonian function, $\{ \ ,\, \}$ denotes the canonical Poisson bracket, while the Hermitian operator $\widehat{L}_H$ is often called \emph{the Liouvillian}. The KvN equation has been rediscovered in several instances \cite{Wiener,tHooft} and it has been attracting some attention in recent years \cite{Bondar,Ghose,Klein,Mauro,Viennot,Wilczek}. For a broad review of general applications of Koopman operators, see also \cite{Mezic}.

Based on the KvN construction, Sudarshan's theory invoked special superselection rules for physical consistency purposes. In turn, these superselection rules lead to interpretative problems which resulted in Sudarshan's work being  overly criticized \cite{Barcelo,PeTe,Terno,Sudarshan2}. 

\subsection{Koopman-van Hove wavefunctions}
As shown in recent work \cite{BoGBTr}, the standard KvN theory fails to comprise the dynamics of  classical phases and therefore it is somewhat incomplete. Indeed, it is evident that the KvN wavefunction in \eqref{KvNeq} is only defined up to phase functions so that $\Psi$ can be simply chosen to be real-valued. Over the years, phase factors have been suitably added to the standard KvN equation \cite{Bondar,Klein}, which was also related to van Hove's work in prequantization \cite{Sugny,boucher}. More particularly, equivalent variants  of the equation
\begin{equation}\label{KvH1eq}
{\rm i}\hbar\partial_t\Psi=\widehat{L}_H\Psi-(p\partial_{p}H-H)\Psi
\end{equation}
made their first appearance in Kostant's work \cite{Kostant} from 1972  (see also \cite{Gunther}) under the name of ``prequantized Schr\"odinger equation''. One recognizes that the phase term appearing in parenthesis identifies the phase-space expression of the Lagrangian.
Only very recently it was shown \cite{BoGBTr} that this phase factor determines nontrivial contributions to the definition of the Liouville probability density, whose expression reads as follows:
\begin{equation}\label{rhomomap1}
\rho=|\Psi|^2 +\partial_{p}( p |\Psi|^2)  + {\rm i} \hbar\{\Psi,\bar\Psi\}
\,,
\end{equation}
where the bar symbol is used to denote complex conjugation.

This result was found by applying standard momentum map methods  to van Hove's prequantization theory. {Inspired by Kirillov \cite{Kirillov}}, in \cite{BoGBTr} the resulting construction was referred to as the \emph{Koopman-van Hove} (KvH) formulation of classical mechanics. Entirely based on prequantization, this theory allows the systematic application of geometric quantization \cite{Ko1970}. Then, a hybrid quantum--classical theory was found in \cite{BoGBTr} by starting with the KvH equation for a two-particle wavefunction $\Psi(q,p,x,s)$ and then quantizing one of the two particles by standard methods. This process yields an equation for a hybrid quantum--classical wavefunction $\Upsilon(q,p,x)$, where $x$ denotes the quantum coordinate.

As pointed out in \cite{BoGBTr}, this partial quantization procedure leads to a Hamiltonian hybrid theory in which both quantum and classical pure states are lost. We recall that classical pure states are defined as extreme points of the convex set of classical probability densities \cite{ChernoffMarsden} and these are given by delta-like Klimontovich distributions. Unlike quantum pure states, which may be \emph{entangled} and not factorizable, classical pure states are completely factorizable. The absence of classical pure states in the general case of hybrid dynamics raises questions about the nature of Hamiltonian trajectories in quantum--classical coupling. Indeed, classical motion is given by a Hamiltonian flow producing characteristic curves representing particle trajectories and thus one is led to ask whether a Hamiltonian flow can still be identified in hybrid dynamics. In this paper, we address this question by extending the Lagrangian (or \emph{Bohmian}) trajectories from quantum hydrodynamics to hybrid quantum--classical systems. To this purpose, we shall exploit the geometric structure of the Madelung transform.

Another question emerging in the context of hybrid quantum--classical dynamics concerns the existence of a continuity equation for the hybrid density, which could then be used to define a hybrid current extending the probability current from standard quantum theory. This is the second question addressed in this paper, which exploits methods from Geometric Mechanics \cite{Sternberg2,HoScSt09,MaRa} to present the explicit hybrid continuity equation in terms of its underlying Hamiltonian structure. In turn, the existence of a continuity equation leads to the question whether the sign of the hybrid density is preserved in time. Here, we shall present an infinite family of hybrid systems for which this is indeed the case.

\subsection{Madelung transform in quantum mechanics}
This paper,   uses the polar form of the wavefunction in order to characterize the Madelung formulation of hybrid quantum--classical dynamics. This work is inspired by  the Madelung-Bohm  hydrodynamic formulation of quantum mechanics \cite{Madelung,Bohm},  whose geometric features were recently revived in \cite{KhMiMo2017}. In order to obtain his equations of quantum hydrodynamics, Madelung replaced the polar form $\psi(x,t)=R(x,t)e^{-{\rm i}S(x,t)/\hbar}$ of the wavefunction into Schr\"odinger's equation ${\rm i}\hbar\partial_t\psi=-m^{-1}\hbar^{2}\Delta\psi/2+V\psi$. This operation yields the following PDE system
\begin{align}\label{MadelungEqs}
&\frac{\partial  { S}}{\partial t}+\frac{|\nabla { S}|^2}{2m}-\frac{\hbar^2}{2m}\frac{\Delta  R}{ R}+ V= 0\,,
\\
&\frac{\partial  R}{\partial t}+\frac{1}{2mR}\operatorname{div}( R^2\nabla {S})= 0\,.
\label{MadelungEqsbis}
\end{align}
The second equation yields the well-known continuity equation for the probability density $D=R^2$. Madelung realized that defining the velocity vector field 
\[
v=\frac{\nabla S}m
\] 
casts the above system into a set of hydrodynamic equations as follows:
\begin{align}\label{MadelungEqs2}
&\frac{\partial v}{\partial t}+ {v}\cdot\nabla v=-\frac1m\nabla\bigg(  V +\frac{\hbar^2}{2m}   \frac{\Delta  \sqrt{D}}{ \sqrt{D}}\bigg)\,,
\qquad\qquad\qquad
\frac{\partial  D}{\partial t}+\operatorname{div}( Dv)= 0\,.
\end{align}
Madelung's equations were the point of departure for Bohm's interpretation of quantum dynamics \cite{Bohm}. Following  previous ideas by de Broglie \cite{DeBroglie}, Bohm interpreted the integral curves of the velocity vector field $v(t,x)$ as the genuine trajectories in space of the physical quantum particle. In this picture, particles are carried by a {\it pilot wave} transporting probability with a velocity $v$ which itself changes in time according to the first equation in \eqref{MadelungEqs2}. Bohmian trajectories, however, are not exactly point particle trajectories: rather, they are trajectories in a fluid Lagrangian sense. More specifically, if the fluid label ${x}_0$ is mapped to its current position ${x}_t$ in terms of a smooth Lagrangian path $\chi$, one writes ${x}_t=\chi(t, x_0)$ and $\chi(t,\cdot)$ is identified with a time-dependent diffeomorphism of the physical space $M$, that is $\chi(t,\cdot)\in\operatorname{Diff}(M)$. Then, Bohmian trajectories are fluid paths satisfying the reconstruction relation 
\beq\label{QuantumPaths}
\partial_t\chi(t, x)=v(t,\chi(t,x))
\,.
\eeq 
 While Bohmian mechanics and pilot wave theory have raised several fundamental interpretative questions, in this paper we shall not dwell upon these issues. The scope of this work is instead to extend the concept of Bohmian trajectories to hybrid quantum--classical systems and exploit the Madelung transform to draw conclusions about the dynamics of the joint quantum--classical  density.

\subsection{Momentum maps and Madelung equations\label{sec:MadelungMomap}}
In this work, we shall follow a geometric approach combining the geometric setting of the quantum Madelung transform with the prequantization theory of  van Hove \cite{VanHove} and Kostant \cite{Ko1970}. Indeed, both these constructions share momentum map structures which will serve as a unifying framework  to describe quantum--classical coupling. The momentum maps appearing in prequantization will be discussed in the next section, while those emerging in quantum hydrodynamics have recently been exploited in the context of quantum chemistry  \cite{FoHoTr}. 

The Madelung momentum map takes the quantum Hilbert space $L^2(M)$ into the dual of the semidirect-product Lie algebra $\mathfrak{X}(M)\,\circledS\,\mathcal{F}(M)$, where $\mathfrak{X}(M)$ denotes the Lie algebra of vector fields on $M$ and $\mathcal{F}(M)$ the space of real valued functions on $M$. More explicitly, the Madelung momentum map $J:L^2(M)\to\mathfrak{X}^*(M)\times{\rm Den}(M)$  is given by
\beq
J(\psi)=\big(\hbar\operatorname{Im}(\psi^*\nabla\psi),|\psi|^2\big)=(mDv,D)
\,.
\label{QuantumMadelungMomap}
\eeq
Here, $\mathfrak{X}^*(M)$ denotes the dual space of  $\mathfrak{X}(M)$, while ${\rm Den}(M)$ denotes the space of densities on $M$.
Upon considering the standard symplectic form $\Omega(\psi_1, \psi_2)= 2\hbar \operatorname{Im} \int_M \!\bar{\psi}_1\psi_2\,\mu$ (here, $\mu$ is the volume form on physical space $M$),  the above momentum map is generated by a unitary representation of the semidirect-product group $\operatorname{Diff}(M)\,\circledS\,\mathcal{F}(M,S^1)$ which reads as follows:
\beq\label{QuantumMadelungRep}
\psi\mapsto \frac1{\sqrt{\operatorname{Jac}({\chi})}}\ \chi_*({e^{-{\rm i}\varphi/\hbar}}\psi)
\,,
\eeq  
with $(\chi,\varphi)\in \operatorname{Diff}(M)\,\circledS\,\mathcal{F}(M,S^1)$. Here, 
$\mathcal{F}(N_1,N_2)$ generally
denotes the space of mappings from the manifold $N_1$ to the manifold $N_2$, so that $\mathcal{F}(M,S^1)$ denotes the space of $S^1$-valued functions on $M$. Moreover, $\operatorname{Diff}(M)$ denotes the group of diffeomorphisms of $M$ while $\operatorname{Jac}({\chi})$ denotes the Jacobian determinant of $\chi$, and $\chi_*$ denotes the push-forward. Upon denoting composition by $\circ$, one writes $\chi_*(e^{-{\rm i}\varphi/\hbar}\psi)=(e^{-{\rm i}\varphi/\hbar}\psi)\circ\chi^{-1}$. The representation \eqref{QuantumMadelungRep} is typically constructed by identifying the Hilbert space $L^2(M)$ with the space of half-densities \cite{BatesWeinstein,FoHoTr}, although here we shall not discuss this particular aspect. More importantly, throughout this paper we shall assume that the elements in $\operatorname{Diff}(M)\,\circledS\,\mathcal{F}(M,S^1)$ have sufficient regularity to ensure that this group is an infinite-dimensional manifold and a topological group with smooth right translation. Also we assume appropriate restrictions of the domain of the action \eqref{QuantumMadelungRep} and the momentum map \eqref{QuantumMadelungMomap}, so that all the operations are well-defined.

The fact that \eqref{QuantumMadelungMomap} identifies a momentum map for the unitary representation \eqref{QuantumMadelungRep} is a direct verification that makes use of the infinitesimal generator corresponding to \eqref{QuantumMadelungRep}, that is
\beq
\label{MadelungInfGen1}
\psi\mapsto -{\rm i}\hbar^{-1}\alpha\psi-u\cdot\nabla\psi-\frac12(\operatorname{div} u)\psi
\,,
\eeq
with $(u,\alpha)\in \mathfrak{X}(M)\,\circledS\,\mathcal{F}(M)$.
In this paper, we shall exploit the Madelung momentum map \eqref{QuantumMadelungMomap} to present the geometric structure of the Madelung equations for hybrid quantum--classical systems. As mentioned previously, these will be described in terms of a hybrid wavefunction $\Upsilon(q,p,x)$, whose polar form will be used to define Bohmian trajectories in the context of hybrid systems.

\subsection{Outline and results}

In Section \ref{sec:KvH}, the Koopman-van Hove formulation of classical mechanics \cite{BoGBTr} is reviewed, along with its underlying geometric structure in terms of strict contact transformations, that is, connection-preserving automorphisms of the prequantum circle bundle $T^*Q\times S^1 \to T^*Q$. This treatment is essentially equivalent to that presented by Kostant \cite{Kostant} in the early 70's. Following \cite{ILM}, we show how the group of strict contact diffeomorphisms is isomorphic to a central extension of the symplectic diffeomorphism group by $S^1$, whose Lie algebra identifies the Poisson algebra of Hamiltonian functions on the classical phase-space $T^*Q$. In Section \ref{sec:rhomomap} we review recent work \cite{BoGBTr} to show how the KvH formulation of classical mechanics produces the classical Liouville equation. This connection is established by a momentum map associated to the unitary action of strict contact diffeomorphisms on the sections of the prequantum bundle, which are here identified with complex wavefunctions on the classical phase-space. In Section \ref{sec:KvHMadelung}, the Madelung transform is applied to the KvH equation \eqref{KvH1eq} to show how the classical phase is naturally incorporated.

Section \ref{sec:CQeq} presents the mathematical setting of the hybrid wave equation for quantum--classical dynamics. The hybrid wavefunction on the hybrid coordinate space $\Gamma=T^*Q\times M$ (here, $M$ is the quantum configuration space) undergoes unitary evolution, whose Hermitian generator is called \emph{hybrid Liouvillian}. The algebraic study of hybrid Liouvillian operators is presented in Section \ref{sec:algebra}, along with a remarkable identity relating commutators and Poisson brackets. In the same section, hybrid Liouvillians are shown to be equivariant under both quantum unitary transformations and classical strict contact transformations. The same long sought equivariance properties \cite{boucher} are shared by a  hybrid density operator extending the quantum density matrix to quantum--classical dynamics, as shown in Section \ref{sec:hybden}. While  the density matrix of the quantum subsystem is positive-definite by construction in all cases, the hybrid density operator is generally unsigned and thus the sign of the classical Liouville density may require a case-by-case study. 

Classical and quantum pure states are shown to be both lost in the general case of hybrid dynamics thereby leading to questions about the existence of trajectories in the case of quantum--classical coupling. Section \ref{sec:HybMad} addresses these questions by applying the Madelung transform to the hybrid wave equation, thereby leading to the identification of hybrid quantum--classical Bohmian trajectories and their generating vector field in Section \ref{sec:Bohmian}. In the presence of a quantum--classical interaction potential, the symplectic form on the classical phase-space is not preserved by the hybrid flow and Section \ref{sec:sympform} characterizes explicitly the nontrivial dynamics of the Poincar\'e integral on the hybrid coordinate space $\Gamma$. Nevertheless, the classical phase-space components of the hybrid Bohmian trajectories identify a Hamiltonian flow parameterized by the quantum coordinate and this flow is associated to the dynamics of the classical subsystem. Also, the Hamiltonian structure of the hybrid Madelung equations is  presented  in Section \ref{Ham_var}.

In Section \ref{Sec_probability}, we consider the geometric structure of the joint quantum--classical density on the hybrid coordinate space $\Gamma$. This hybrid density is found to be a momentum map in Section \ref{sec:Dmomap} and this ensures preservation of its sign in the special case when the quantum kinetic energy is absent in the hybrid Hamiltonian. Section \ref{continuity_calD} presents the continuity equation for the hybrid density, thereby leading to the identification of a hybrid quantum--classical current analogue to the probability current in quantum mechanics. The hybrid continuity equation is then shown to possess a Lie-Poisson Hamiltonian structure in Section \ref{sec:bracket}. The paper closes with Section \ref{sec:Hamclass}, which identifies an infinite family of hybrid Hamiltonians producing a quantum--classical dynamics that preserve the sign of the classical probability density.


\section{Koopman-van Hove classical mechanics\label{sec:KvH}}

In this Introduction, we shall review the KvH theory developed in \cite{BoGBTr} and present some of its features, along with its Madelung representation.

\subsection{The Koopman-van Hove equation}

Let $Q$ be the configuration manifold of the classical mechanical system and $T^*Q$ its phase space, given by the  cotangent bundle of $Q$. We assume that the manifold $Q$ is connected. We shall denote by $z\in T^*Q$ an element of the phase space, and write $z=(q^i,p_i)$ in local coordinates. The phase space is canonically endowed with the one-form ${\mathcal{A}}= p_i{\rm d}q^i$ and the symplectic form $\omega=-{\rm d}{\mathcal{A}}= {\rm d} q^i\wedge {\rm d}p_i$, where $\de$ denotes the exterior derivative. For later purpose, it is also convenient to consider the trivial principal circle bundle 
\begin{equation}\label{prequantumbundle}
T^*Q\times S^1\rightarrow T^*Q
\end{equation}
(known as \emph{prequantum bundle}) in such a way that ${\mathcal{A}}$ identifies a principal connection ${\mathcal{A}}+ \de s$ with curvature given by (minus) the symplectic form $\omega$.

A classical wave function $\Psi$ is an element of the complex Hilbert space 
\[
\mathscr{H}_{\scriptscriptstyle C}=L^2(T^*Q)
\]
with standard Hermitian inner product
\[
\langle \Psi_1| \Psi_2\rangle= \int_{T^*Q\!} \bar{\Psi}_1(z)\Psi_2(z)\,\Lambda\qquad
\text{with}
\qquad
\Psi_1,\Psi_2\in \mathscr{H}_{\scriptscriptstyle C},
\]
defined in terms of the Liouville volume form $\Lambda=(-1)^{n(n-1)/2}\omega^n/n!$ (the multiplicative factor is such that in local coordinates  one has $\Lambda= \de q^1\wedge ... \wedge \de q^n \wedge \de p_1 \wedge ... \wedge\de p_n$).
The corresponding real-valued pairing and symplectic form on $\mathscr{H}_{\scriptscriptstyle C}$ are given by
\begin{equation}\label{inner_symplectic}
\langle \Psi_1, \Psi_2\rangle= \operatorname{Re} \int_{T^*Q\!}  \bar{\Psi}_1(z)\Psi_2(z)\,\Lambda
\qquad\text{and}\qquad\Omega(\Psi_1, \Psi_2)= 2\hbar \operatorname{Im} \int_{T^*Q\!} \bar{\Psi}_1(z)\Psi_2(z)\,\Lambda
\,.
\end{equation}

Given a classical Hamiltonian function $H\in {\cal F}(T^*Q)$, the \textit{KvH equation for classical wavefunctions} was presented in \cite{BoGBTr,Gunther,Sugny,Klein,Kostant} and it reads
\begin{equation}\label{KvH_eq}
{\rm i}\hbar\partial_t\Psi={\rm i}\hbar \{ H,\Psi\}-({\mathcal{A}}\!\cdot\! X_H-H)\Psi\,.
\end{equation}
Here, $X_H$ is the Hamiltonian vector field associated to $H$, i.e. $\mathbf{i}_{X_H}\omega={\rm d}H$, and $\{H,K\}=\omega(X_H, X_K)$ is the canonical Poisson bracket, extended in \eqref{KvH_eq} to $\mathbb{C}$-valued functions by $\mathbb{C}$-linearity. Note that ${\mathcal{A}}\!\cdot\! X_H= p_i \partial_{p_i} H$ in local coordinates, thereby recovering the KvH equation \eqref{KvH1eq} for a  one-dimensional configuration manifold $Q$. Then, ${\mathcal{A}}\cdot X_H-H$ identifies the Lagrangian associated to $H$ and the entire right hand side of \eqref{KvH_eq} defines the \emph{covariant Liouvillian operator}
\beq\label{preqop}
\widehat{\cal L}_H={\rm i}\hbar \{H,\ \}-({\mathcal{A}}\!\cdot\! X_H-H)
\,.
\eeq
Also known as \emph{prequantum operator}, this is easily seen to be an unbounded Hermitian operator on $\mathscr{H}_{\scriptscriptstyle C}$. As a consequence, the KvH equation \eqref{KvH_eq} comprises a Hamiltonian system with respect to the symplectic form \eqref{inner_symplectic} and Hamiltonian functional
\[
h(\Psi)= \int_{T^*Q}\bar\Psi \widehat{\cal L}_H\Psi\,\Lambda\,.
\]
The correspondence $H\mapsto\widehat{\cal L}_H$ satisfies $[\widehat{\cal L}_H,\widehat{\cal L}_F]={\rm i}\hbar\widehat{\cal L}_{\{H,F\}}$, for all $H,F\in C^\infty(T^*Q)$.
Hence,  on its domain, the operator $\Psi\mapsto -{\rm i}{\hbar }^{-1}\widehat{\cal L}_H\Psi$ defines a  skew-Hermitian (or, equivalently, symplectic) left representation of the Lie algebra $({\cal F}(T^*Q),\{\ ,\,  \})$  on the classical Hilbert space $\mathscr{H}_{\scriptscriptstyle C}$. Note that, unlike the map $H \mapsto \widehat{L}_H={\rm i}\hbar\{H,\ \}$ in \eqref{KvNeq}, the correspondence $H\mapsto\widehat{\cal L}_H$ is now injective, i.e. $ \widehat{\cal L}_H=\widehat{\cal L}_F\iff H=F$.

\subsection{The group of strict contact diffeomorphisms}

In this Section, we shall assume that  the first cohomology group ${\sf H}^1(T^*Q, \mathbb{R})=0$ (or, equivalently, ${\sf H}^1(Q, \mathbb{R})=0$). Under this assumption, we shall follow van Hove \cite{VanHove} and show that the operator $-{\rm i}{\hbar }^{-1}\widehat{\cal L}_H$ integrates to a unitary left representation, whose corresponding group is characterized below. 

As a preliminary step,  given the trivial circle bundle \eqref{prequantumbundle}, we consider its automorphism group given by the semidirect product $\operatorname{Diff}(T^*Q)\,\circledS\, \mathcal{F}(T^*Q, S^1)$. As explained in Section \ref{sec:MadelungMomap}, this group carries a natural  unitary representation on the classical Hilbert space $\mathscr{H}_{\scriptscriptstyle C}$, which reads
\beq\label{sdp-action}
\Psi\mapsto \frac1{\sqrt{\operatorname{Jac}(\eta)}}\ \eta_*({e^{-i\varphi/\hbar}}\Psi)
\,,
\eeq  
with $(\eta,\varphi)\in\operatorname{Diff}(T^*Q)\,\circledS\, \mathcal{F}(T^*Q, S^1)$. This is essentially the same representation as in \eqref{QuantumMadelungRep}, upon replacing the quantum configuration space $M$ with the classical phase space $T^*Q$.
Likewise, the  infinitesimal generator corresponding to the unitary action \eqref{sdp-action} is again the analogue of \eqref{MadelungInfGen1} and one gets
\beq\label{sdp-infaction}
\Psi\mapsto -{\rm i}\hbar^{-1}\nu\Psi- X\cdot\nabla \Psi-\frac12(\operatorname{div} X)\Psi
\,,
\eeq
where $(X,\nu)\in \mathfrak{X}(T^*Q)\,\circledS\,\mathcal{F}(T^*Q)$ 
and $\operatorname{div}$ is the divergence with respect to the Liouville volume form $\Lambda$ on $T^*Q$.
As already anticipated, the representation \eqref{sdp-action} and its infinitesimal generator \eqref{sdp-infaction} will be of fundamental importance in later sections.

A relevant subgroup of the semidirect product $\operatorname{Diff}(T^*Q)\,\circledS\, \mathcal{F}(T^*Q, S^1)$ is given by those transformations preserving the connection one-form $\mathcal{A}+\de s$ on $T^*Q\times S^1$, that is the group $\operatorname{Aut}_{\cal A}(T^*Q\times S^1)$ of connection-preserving automorphisms of the principal bundle \eqref{prequantumbundle}. More explicitly, one has 
\begin{equation}\label{stricts}
\operatorname{Aut}_{\cal A}(T^*Q\times S^1):=
\left\{(\eta,e^{\rm i\varphi})\in\operatorname{Diff}(T^*Q)\,\circledS\, \mathcal{F}(T^*Q, S^1)\ \Big|\ \eta^*\mathcal{A}+\de\varphi=\mathcal{A} \right\}
,
\end{equation}
where $\eta^*$ denotes pullback.
The above transformations were studied extensively  in van Hove's thesis \cite{VanHove} and are known as forming the group of \emph{strict contact diffeomorphisms} \cite{Grey}. This group is related to the more familiar group $\operatorname{Diff}_\omega(T^*Q)$ of symplectic diffeomorphisms (canonical transformations) and this relation will be discussed in detail in the next section. For the moment, we simply notice that the relation $\eta^*\mathcal{A}+\de\varphi=\mathcal{A}$ implies
\beq\label{Aut-relations}
\eta^*(\de{\cal A})=0
\,,\qquad\qquad
\varphi(z)=\theta+\int_{z_0}^{z}({\cal A}-\eta^*{\cal A})
\,,
\eeq
so that $\eta\in \operatorname{Diff}_\omega(T^*Q)$ and  $\varphi$ is determined up to a constant phase $\theta=\varphi(z_0)$. Since ${\sf H}^1(T^*Q, \mathbb{R})=0$, the line integral above does not depend on the curve connecting  $z_0$ to $z$.
As a subgroup of the semidirect product $\operatorname{Diff}(T^*Q)\,\circledS\, \mathcal{F}(T^*Q, S^1)$, the group $\operatorname{Aut}_{\cal A}(T^*Q\times S^1)$ inherits from \eqref{sdp-action} a unitary representation, which is obtained essentially by replacing \eqref{Aut-relations} in \eqref{sdp-action}. As we shall show in the next section, the operator ${\rm i}{\hbar }^{-1}\widehat{\cal L}_H$ emerges as the infinitesimal generator of this representation.

The relations \eqref{Aut-relations} have an immediate correspondent at the level of the Lie algebra $\mathfrak{aut}_{\cal A}(T^*Q\times S^1)$, which can be initially defined  by using Lie derivatives as
\begin{equation*}
\mathfrak{aut}_{\cal A}(T^*Q\times S^1)=
\Big\{(X,{\nu})\in\mathfrak{X}(T^*Q)\,\circledS\, \mathcal{F}(T^*Q)\ \Big|\ \pounds_X\mathcal{A}+\de\nu=0  \Big\}
.
\end{equation*}
We notice that the relation $\pounds_X\mathcal{A}+\de\nu=0$ implies $\pounds_X(\de\mathcal{A})=0$ thereby identifying a Hamiltonian vector field $X=X_H$, for some $H\in \mathcal{F}(T^*Q)$. In turn,  Cartan's magic formula yields $\pounds_X\mathcal{A}=\de(\mathcal{A} \cdot X_H-H)$, so that $\de(\nu+ \mathcal{A} \cdot X_H-H)=0$ and eventually one is left with
\[
X=X_H
\,,\qquad\qquad
\nu=H- \mathcal{A} \cdot X_H
\,,
\]
where an integration constant has been absorbed into the Hamiltonian $H$. For later purpose, here we introduce the notation 
\begin{equation}\label{indextheta}
F_{\mathcal{A}} = {\mathcal{A}}\!\cdot \!X_F-F\,,
\end{equation}
for the Lagrangian associated to the function $F$.
It is now evident that any smooth Hamiltonian function $H\in \mathcal{F} (T^*Q)$ determines a Lie algebra element in $(X,\nu)\in\mathfrak{aut}_{\cal A}(T^*Q\times S^1)$ via the map
\begin{equation}\label{H_X_nu}
H \in \mathcal{F} (T^*Q) \rightarrow (X_H, - H_ \mathcal{A}) \in \mathfrak{aut}_{\cal A}(T^*Q\times S^1).
\end{equation}
Analogously, any pair $(\eta,\theta)\in \operatorname{Diff}_\omega(T^*Q)\times S^1$ determines a group element in $\operatorname{Aut}_{\cal A}(T^*Q\times S^1)$. This picture can be given an equivalent and more convenient geometric structure in terms of central extensions.

\subsection{A central extension of symplectic diffeomorphisms}\label{central_ext_sym}

Given the group $\operatorname{Diff}_\omega(T^*Q)$ of symplectic diffeomorphisms, let us fix a point $z_0\in T^*Q$ and introduce the group 2-cocycle
\begin{equation}\label{groupcocycle}
B_{z_0}(\eta_1,\eta_2) := \int_{z_0}^{\eta_2(z_0)}\left({\mathcal{A}} -\eta_1^*{\mathcal{A}}\right),
\end{equation}
given by the line integral of the one form ${\mathcal{A}} -\eta_1^*{\mathcal{A}}$ along a path connecting $z_0$ to $\eta_2(z_0)$. As discussed previously,  ${\mathcal{A}} -\eta_1^*{\mathcal{A}}$ is exact since  ${\sf H}^1(T^*Q, \mathbb{R})=0$. Thus, the relation ${\mathcal{A}} -\eta_1^*{\mathcal{A}}=\de\varphi_1$ leads to $B_{z_0}(\eta_1,\eta_2) =\eta_2^* \varphi_1(z_0)- \varphi_1(z_0)$ whose value is independent of the integration path.
Also, as reported in \cite{ILM}, the cohomology class of $B_{z_0}$ is independent of both choices for  the point $z_0$ and  the $1$-form ${\mathcal{A}}$, where we recall $-{\rm d}{\mathcal{A}}=\omega$.

Here, we shall use the group 2-cocycle \eqref{groupcocycle} to express the group \eqref{stricts} of strict contact diffeomorphisms as a convenient central extension of the group $\operatorname{Diff}_\omega(T^*Q)$ of symplectic diffeomorphisms by the circle group $S^1$. In particular, we use the group 2-cocycle \eqref{groupcocycle} to construct the following central extension:
\begin{align}\label{scont}
\widehat{\operatorname{Diff}}_\omega(T^*Q)=\operatorname{Diff}_\omega(T^*Q)\times _{\scriptscriptstyle B_{z_0}\!} S^1
\,,
\end{align}
 endowed with the  group product structure \cite{GBTr,ILM}
\begin{equation}\label{mult_scont}
(\eta_1,e^{{\rm i}\theta_1}) (\eta_2,e^{{\rm i}\theta_2})=(\eta_1\circ\eta_2,e^{{\rm i}\theta_1+{\rm i }\theta_2 + {\rm i}B_{z_0}(\eta_1,\eta_2)}),\quad
\end{equation}
where   $z_0\in T^*Q$ is some  fixed point.

\noindent
At this stage, since the prequantum bundle  \eqref{prequantumbundle} is trivial we have the following statement.
\begin{proposition}[\cite{ILM}]\label{jimmy}
Given an element $(\eta, e^{{\rm  i} \theta})\in \widehat{\operatorname{Diff}}_\omega(T^*Q)$, the following mapping defines a group isomorphism $\widehat{\operatorname{Diff}}_\omega(T^*Q)\to\operatorname{Aut}_{\cal A}(T^*Q\times S^1)$:
\beq
(\eta, e^{{\rm  i} \theta})
\mapsto 
\left(\eta,e^{{\rm i}\theta} e^{{\rm i}\!\int_{z_0}^{z}\left({\mathcal{A}} -\eta^*{\mathcal{A}}\right)}\right).
\label{groupisomorphism}
\eeq
The inverse isomorphism $\operatorname{Aut}_{\cal A}(T^*Q\times S^1)\to\widehat{\operatorname{Diff}}_\omega(T^*Q)$ is given  by $(\eta, e^{{\rm i}\varphi})
\mapsto (\eta, e^{{\rm i}\varphi(z_0)})
$.
\end{proposition}

Once the group structure of the central extension \eqref{scont} is characterized, one can find its corresponding Lie algebra structure. The latter is given by the central extension of the Lie algebra of symplectic (hence Hamiltonian) vector fields $\mathfrak{X}_\omega(T^*Q)$ denoted
\beq\label{LAext}
\widehat{\mathfrak{X}}_\omega(T^*Q):= \mathfrak{X}_\omega(T^*Q)\times _{\scriptscriptstyle C_{z_0}\!}\mathbb{R}
\,.
\eeq
Here, $C_{z_0}$ is the Lie algebra 2-cocyle associated to $B_{z_0}$ so that
\[
\big[(X_H, \kappa ),(X_F, \gamma )\big]=\big(X_{\{H,F\}},C_{z_0}(X_H, X_F)\big),
\]
for all $(X_H, \kappa ),(X_F, \gamma ) \in \widehat{\mathfrak{X}}_\omega(T^*Q)$.  The Lie algebra 2-cocyle is constructed as follows: given two Hamiltonian vector fields $X_H,X_F\in{\mathfrak{X}}_\omega(T^*Q)$   whose flows are denoted respectively by $ \eta _1(t)$ and $ \eta _2(s)$,  we have
\begin{equation}\label{relation_B_C} 
C_{z_0}(X_H, X_F) = \left. \frac{d}{d s}\right|_{s=0}  \left. \frac{d}{d t}\right|_{t=0} \left(  B_{z_0} ( \eta_1 (t) ^{-1} , \eta_2 (s))  - B_{z_0} ( \eta_2 (s) , \eta_1 (t)^{-1} ) \right).
\end{equation} 
After a direct calculation using the notation introduced in \eqref{indextheta}, we obtain \cite{GBTr}
\[
C_{z_0}(X_H, X_F)={\mathcal{A}}\!\cdot \!X_{\{F,H\}}(z_0)-\{F,H\}(z_0)=\{F,H\}_{\mathcal{A}}(z_0)
\,.
\]
Then, as the Lie algebra structure of \eqref{LAext} is now characterized, proposition \ref{jimmy} leads naturally to the following result which is the analogue of \eqref{H_X_nu}:
\begin{proposition}[\cite{GBTr}]\label{LAprop}
Given a point $z_0\in T^*Q$, the following mapping
\begin{equation}\label{LAisomorphism} 
H \in \mathcal{F}(T^*Q) \mapsto (X_H, -H_{\mathcal{A}}(z_0)) \in \widehat{\mathfrak{X}}_\omega (T^*Q)
\end{equation} 
is a Lie algebra isomorphism.
The inverse isomorphism $ \widehat{\mathfrak{X}}_\omega (T^*Q) \to   \mathcal{F}(T^*Q)$ is given by ${(X, a)  \mapsto  H}$, where $H$ is the unique Hamiltonian of $X$ with $H(z_0)= \mathcal{A} \!\cdot \!X (z_0) + a$.
\end{proposition}
\paragraph{Proof.} Taking the derivative of the group isomorphism \eqref{groupisomorphism} at the identity yields Lie algebra isomorphism
\begin{equation}\label{Lie_algebra_isom_aut} 
(X_H, \kappa ) \in \widehat{\mathfrak{X}}_\omega (T^*Q) \mapsto (X_H, \kappa  + H_ \mathcal{A}(z_0) -  H_ \mathcal{A}) \in \mathfrak{aut}_{\cal A}(T^*Q\times S^1)
\,,
\end{equation} 
whose inverse is simply given by $(X_H, \nu  ) \in \mathfrak{aut}_{\cal A}(T^*Q\times S^1)\mapsto (X_H, \nu  (z_0)) \in \widehat{\mathfrak{X}}_\omega (T^*Q) $.
By composing with the isomorphism \eqref{H_X_nu} we do get \eqref{LAisomorphism}. The Lie algebra isomorphism property of \eqref{LAisomorphism}  
\beq\label{LAisomorphism2}
\{H,F\} \in \mathcal{F}(T^*Q) \mapsto\big(X_{\{H,F\}},-\{H,F\}_{\mathcal{A}}(z_0)\big) \in \widehat{\mathfrak{X}}_\omega (T^*Q)
\eeq
 follows from a direct computation.  $\qquad\qquad\blacksquare$

\subsection{The van Hove representation and the Liouville density\label{sec:rhomomap}}

At this stage, we can rewrite the unitary (left) representation of the group $\operatorname{Aut}_{\cal A}(T^*Q\times S^1)$ on the classical Hilbert space  as an action of the central extension $\widehat{\operatorname{Diff}}_\omega(T^*Q)$. Indeed, if $(\eta, e^{{\rm i}\theta})\in \widehat{\operatorname{Diff}}_\omega(T^*Q)$, its representation on $\mathscr{H}_{\scriptscriptstyle C}$ is obtained by  using \eqref{groupisomorphism} in \eqref{sdp-action}. We have
\begin{equation}\label{cl_prop}
\left[U_{(\eta, e^{{\rm i}\theta})}\Psi \right](z)=\Psi( \eta^{-1}(z))\exp\left[{-\frac{\rm i}{\hbar}\left(\theta+\int_{\eta(z_0)}^{z}(\eta_*{\mathcal{A}} - {\mathcal{A}})\right)}\right]\,.
\end{equation}
The property $U_{(\eta_1, e^{{\rm i}\theta_1})} \circ U_{(\eta_2, e^{{\rm i}\theta_2})}= U_{(\eta_1, e^{{\rm i}\theta_1}) (\eta_2,e^{{\rm i}\theta_2})}$ can be directly  verified by using \eqref{mult_scont}. This representation made its first appearance in \cite{VanHove} as a unitary representation  of the group \eqref{stricts} and here we shall call it the {\it van Hove representation}. 

Analogously, the Lie algebra representation of $(X_H,\kappa)\in \widehat{\mathfrak{X}}_\omega (T^*Q)$ associated to \eqref{cl_prop} is computed as
\begin{align}\nonumber
u_{(X, \kappa)}\Psi = \frac{\de}{\de\epsilon}\, U_{(\eta_\epsilon, e^{{\rm i}\theta_\epsilon})}\Psi\,\bigg|_{\epsilon=0}=&\, -  {\rm i}\hbar ^{-1}( \kappa+H_{\mathcal{A}}(z_0)-H_{\mathcal{A}} )\Psi   - X_H \cdot\nabla\Psi 
\\
=&\, 
-{\rm i}\hbar ^{-1}( H- {\mathcal{A}} \! \cdot \!X_H )\Psi   -  X_H \cdot\nabla\Psi
\label{LA_action}\,,
\end{align}
where $(\eta_\epsilon, e^{{\rm i}\theta_\epsilon})\in \widehat{\operatorname{Diff}}_\omega(T^*Q)$ is a path tangent to $(X_H,\kappa)$ at $(id,1)$,  $H\in \mathcal{F}(T^*Q)$ is an arbitrary function, and $H_{\mathcal{A}}$ is obtained from the notation \eqref{indextheta}. Then, the Lie algebra representation \eqref{LA_action} coincides with $-{\rm i}{\hbar}^{-1}\widehat{\cal L}_H$ as claimed previously. 
We emphasize that, since $-{\rm i}{\hbar}^{-1}\widehat{\cal L}_H$ is the infinitesimal generator of the representation \eqref{cl_prop}, the prequantum operator $\widehat{\cal L}_H$ is equivariant with respect to the action of $\widehat{\operatorname{Diff}}_\omega(T^*Q)$, namely
\begin{equation}\label{equivariance_L_H}
U_{(\eta, e^{{\rm i}\theta})}^\dagger \widehat{\cal L}_H U_{(\eta, e^{{\rm i}\theta})}= \widehat{\cal L}_{H\circ\eta}.
\end{equation}
This relation was used in \cite{BoGBTr} to write Koopman-van Hove dynamics in the Heisenberg picture. In the present work, we shall extend this result to the case of hybrid quantum--classical systems; see Section \ref{sec:algebra}.

So far, nothing has been said about how the KvH equation \eqref{KvH_eq} is related to classical mechanics. As shown in \cite{BoGBTr}, this relation is given in terms of a momentum map $\mathscr{H}_{\scriptscriptstyle C}\to\operatorname{Den}(T^*Q)$, where $\operatorname{Den}(T^*Q)$ denotes the space of densities on $T^*Q$. Since the van Hove representation \eqref{cl_prop} is unitary, it is symplectic with respect to the symplectic form \eqref{inner_symplectic} and thus admits a momentum map $\rho(\Psi)$ via the standard formula \cite{Sternberg2,HoScSt09,MaRa}
\begin{equation}\label{MomapFormula}
\langle\rho(\Psi),H\rangle=\frac{1}{2}\Omega\big(-{\rm i}\hbar ^{-1}{\widehat{\cal L}}_H\Psi,\Psi\big)
\,.
\end{equation}
Here, $\langle\ ,\,\rangle$ denotes the duality pairing between $\mathcal{F}(T^*Q)$ and its dual $\operatorname{Den}(T^*Q)$. Throughout this paper, the angle brackets always denote a duality pairing, whose explicit expression may differ depending on the particular vector space under consideration. A direct calculation \cite{BoGBTr} leads to the momentum map
\begin{align}\nonumber
\rho(\Psi)=&\ |\Psi|^2 - \operatorname{div}\!\big( \mathbb{J} {\mathcal{A}} |\Psi|^2\big)  + {\rm i} \hbar\{\Psi,\bar\Psi\}
\\
 =&\ 
 |\Psi|^2 - \operatorname{div}\!\big(\bar \Psi \mathbb{J}( {\mathcal{A}} \Psi + {\rm i} \hbar\nabla\Psi)\big)\,,
\label{KvHmomap}
\end{align}
where the divergence is associated to the Liouville form and where $\mathbb{J}:T^*(T^*Q)\rightarrow T(T^*Q)$ is defined by $\{F,H\}= \langle {\rm d}F, \mathbb{J}({\rm d}H)\rangle$.
 In local coordinates the second term reads $- \operatorname{div}\!\big( \mathbb{J} {\mathcal{A}} |\Psi|^2\big)= \partial_{p_i}(p_i |\Psi|^2)$. This momentum map is formally a Poisson map with respect to the canonical Poisson structure
 \[
\{\!\!\{f,h\}\!\!\}(\Psi)=\frac1{2\hbar}\operatorname{Im}\int_{ T^*Q}\ \overline{\!\frac{\delta f}{\delta \psi}\!}\ \, \frac{\delta h}{\delta \psi}\,\Lambda
 \] 
 on $\mathscr{H}_{\scriptscriptstyle C}$ and the Lie-Poisson structure 
 \[
\{\!\!\{f,h\}\!\!\}(\rho)=\int_{T^*Q}
\rho\left\{\frac{\delta f}{\delta \rho},\frac{\delta h}{\delta \psi}\right\} \Lambda
 \]
 on $\operatorname{Den}(T^*Q)$. Hence, if $\Psi(t)$ is a solution of 
the KvH equation, the density \eqref{KvHmomap} solves the Liouville equation $\partial_t\rho=\{H,\rho\}$.
As remarked in \cite{BoGBTr}, a density of the form \eqref{KvHmomap} is not necessarily positive definite. However, the Liouville equation generates the sign-preserving evolution $\rho(t)=\eta(t)_*\rho_0$, where $\eta(t)$ is the flow of $X_H$, thereby recovering the usual probabilistic interpretation.

Notice that the momentum map \eqref{KvHmomap} yields the following relation for classical expectation values: given a classical observable $A\in \mathcal{F}(T^*Q)$, its expectation value $\langle A\rangle:=\int_{\scriptscriptstyle T^{*\!}Q} A\rho \,\Lambda$ is expressed as
\beq\label{expvalc}
\langle A\rangle = \int_{ T^*Q} \bar\Psi\widehat{\cal L}_A\Psi\,\Lambda
\,,
\eeq
which is different from the usual expressions appearing in quantum theory. 
At this stage, the meaning of the KvH equation \eqref{KvH_eq} is still somewhat obscure and we shall try to shed some new light by applying the Madelung transform.

\subsection{The Madelung transform \label{sec:KvHMadelung}} 

The Madelung transform of the KvH equation \eqref{KvH_eq} is obtained by writing $\Psi$ in polar form
\[
\Psi(t,z)=R(t,z)e^{{\rm i}S(t,z)/\hbar}
\,,
\]
thereby leading to the following equations for $R$ and $S$
\begin{align}\label{KvHMadelung1}
\partial_t S+\{S,H\}=&\ L
\\
\partial_t R+\{R,H\}=&\ 0
\,.
\label{KvHMadelung2}
\end{align}
Here, we have introduced the Lagrangian $L = p_i\partial_{p_i} H-H \in \mathcal{F}(T^*Q)$, or equivalently, using the notation in \eqref{indextheta}, $L:=H_{\mathcal{A}}$.
Thus, while \eqref{KvHMadelung2} recovers the standard Koopman-von Neumann equation for the amplitude $|\Psi|$, the KvH construction comprises also the dynamics \eqref{KvHMadelung1} of the classical phase. Notice that \eqref{KvHMadelung1} is equivalently written as
\begin{equation}\label{classicalphase_evolution}
\frac{\de }{\de t}S(\eta(t,z),t)=L(\eta(t,z))
\end{equation}
where $\eta(t)$ is the flow of $X_H$.  If the right-hand side in \eqref{classicalphase_evolution}  is set to zero, one recovers the phase evolution arising from the Koopman-von Naumann equation \eqref{KvNeq}. However, in the Koopman-van Hove construction under consideration, the Lagrangian function $L={\cal A}\cdot X_H-H$ is retained in the expression \eqref{preqop} of the covariant Liouvillian $\widehat{\cal L}_H$ and equation \eqref{classicalphase_evolution} is solved formally as follows:
\begin{equation}\label{S_formula}
S(z,t)=\int_{t_0}^t\!L(\eta(\tau-t, z))\,{\rm d} \tau+S(\eta(t_0 -t,z),t_0)
\,.
\end{equation}
We remark that, since $\pounds_{X_H}\mathcal{A}= \de H_\mathcal{A}= \de L$, the phase dynamics also produces the relation
\begin{equation}\label{dS_theta}
(\partial_t+\pounds_{X_H})({\rm d} S-{\mathcal{A}})=0\,,
\end{equation}
which is written in terms of the Lie derivative $\pounds_{X_H}={\rm d} \mathbf{i}_{X_H} + \mathbf{i}_{X_H}{\rm d}$. Notice that the relation ${\rm d} S={\mathcal{A}}$ would be preserved in time thereby recovering the KvN prescription $\rho=|\Psi|^2$ via the momentum map \eqref{KvHmomap}. However, as pointed out in \cite{Ilon}, this possibility would  introduce topological singularities which we shall not treat on this occasion. Instead, here  we notice that the relation \eqref{dS_theta} implies
$
\eta(t)^*(\de S(t)-{\mathcal{A}})=\de S_0-{\mathcal{A}}
$,
or, equivalently,
\[
\de (S(t)-\eta(t)_*S_0)= {\mathcal{A}} - \eta(t)_*{\mathcal{A}}\,.
\]
This is the customary relation for generating functions \cite{MaRa} and it is consistent with
\begin{equation}\label{simplecticity}
\frac{\rm d}{{\rm d} t} \eta(t)^* \omega=0
\,,
\end{equation}
which follows directly from the fact that $\eta(t)$ is the flow of $X_H$.

The amplitude equation also retains some interesting features. Indeed, we notice that defining $D=R^2$ yields the Liouville-type equation
\[
\partial_t D+\{D,H\}=0
\,,
\]
which formally allows for the singular solution
\[
D(z,t)=\delta (z-\zeta(t))
\]
where the curve $\zeta(t)\in T^*Q$ satisfies the Hamilton equations ${\rm d}\zeta/{\rm d }t = X_H(\zeta)$. The particle phase along $\zeta(t)$ is deduced from \eqref{classicalphase_evolution} by writing $\zeta(t)=\eta(t,z_0)$ for some $z_0\in T^*Q$ as
\[
S(\zeta(t), t)=\int_{t_0}^t L(\zeta(\tau)){\rm d}\tau + S(\zeta(t_0),t_0).
\]
While this process is only formal (the relation $D=R^2$ prevents $D$ from being a delta function), these relations are somewhat revealing of a finite-dimensional correspondent of KvH theory.

To conclude this section, we present the relation between the momentum map \eqref{KvHmomap} for the classical Liouville equation and the KvH analogue of the hydrodynamic momentum map \eqref{QuantumMadelungMomap} associated to the Madelung transform. This KvH analogue reads 
\begin{equation}\label{Joli}
J(\Psi)=(\hbar\operatorname{Im}(\bar\Psi\nabla\Psi),|\Psi|^2)=:(\sigma,D)
\end{equation} 
and  is associated to the representation \eqref{sdp-action}  of the prequantum bundle automorphisms $\operatorname{Aut}(T^*Q\times S^1)$ on $\mathscr{H}_{\scriptscriptstyle C}$. 
 As we have seen in Section \ref{central_ext_sym}, this representation reduces to \eqref{cl_prop} upon restricting to the subgroup of connection-preserving automorphisms $ \widehat{\operatorname{Diff}}_\omega(T^*Q)\simeq \operatorname{Aut}_{\cal A}(T^*Q\times S^1)\subset \operatorname{Aut}(T^*Q\times S^1)$. Thus, the momentum map $\rho(\Psi)$ in \eqref{KvHmomap} for the classical Liouville equation can be related to $J(\Psi)$ in terms of the dual of the Lie algebra inclusion $\iota: \mathcal{F}(T^*Q) \hookrightarrow \mathfrak{X}(T^*Q)\,\circledS\, \mathcal{F}(T^*Q)$. Here we recall the Lie algebra isomoprhism $\mathcal{F}(T^*Q)\simeq\widehat{\operatorname{Diff}}_\omega(T^*Q)$ from Proposition \ref{LAprop}.  More explicitly, using the notation \eqref{indextheta}, the Lie algebra inclusion is given by 
 \beq
 \iota(H)=(X_H, -H_{\mathcal{A}})
 \label{LAincl}
 \eeq
  and thus its dual map $\iota^*: \mathfrak{X}(T^*Q)^*\times \operatorname{Den}(T^*Q)\rightarrow \operatorname{Den}(T^*Q)$ reads $\iota^*(\sigma, D)= D - \operatorname{div}({\mathbb{J}{\mathcal{A}} D} - \mathbb{J}\sigma)$. Then, as one verifies explicitly, one obtains
$
\iota^*[J(\Psi)]= \rho(\Psi)$
for all $\Psi\in \mathscr{H}_{\scriptscriptstyle C}$.
This indeed provides an important relation between the momentum map \eqref{KvHmomap} for the classical Liouville equation and the momentum map \eqref{Joli} associated to the KvH Madelung transform.

\smallskip
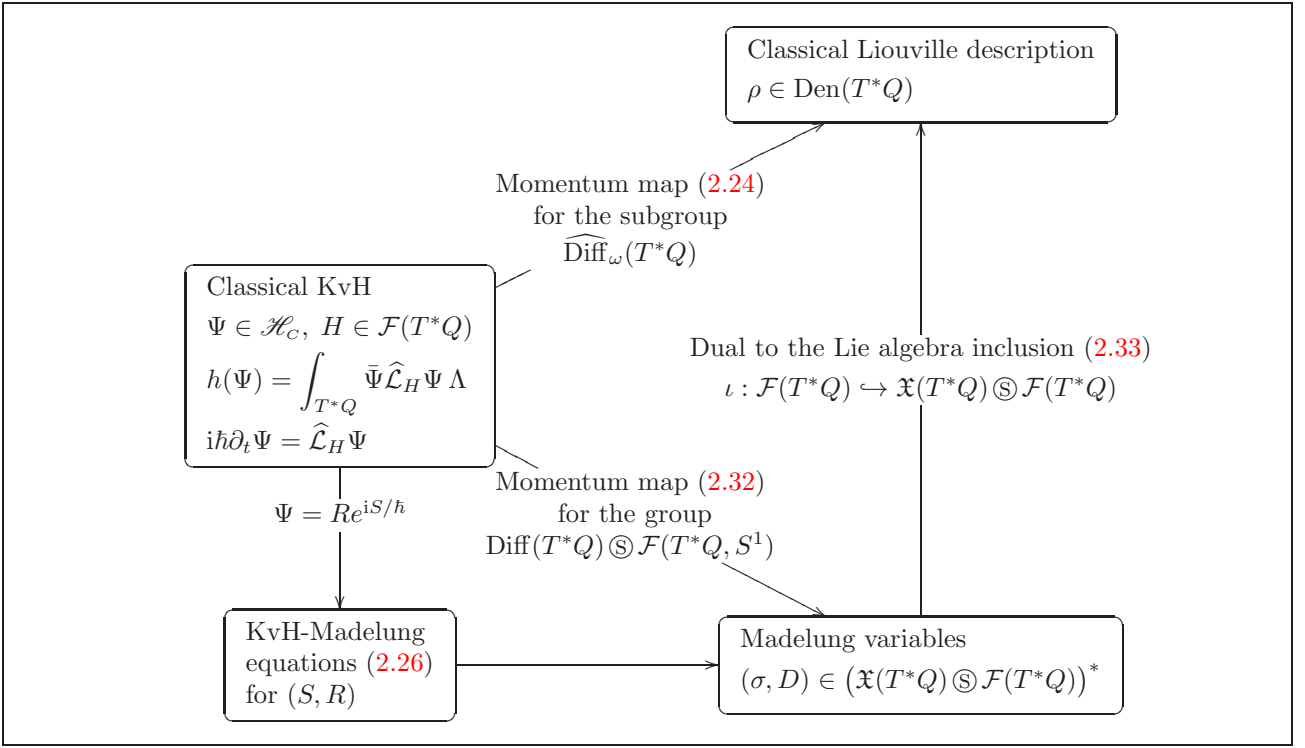
\begin{figure}[h]
\footnotesize\center
\noindent
\begin{framed}
\begin{xy}
\hspace{.9cm}
\xymatrix{
& & &  &*+[F-:<3pt>]{
\begin{array}{l}
\vspace{0.1cm}\text{Classical Liouville description}\\
\vspace{0.1cm} \rho  \in \operatorname{Den}(T^*Q)
\end{array}
} &\\
& & & & &\\
&
*+[F-:<3pt>]{
\begin{array}{l}
\vspace{0.1cm}\text{Classical KvH}\\
\vspace{0.1cm}\Psi \in \mathscr{H}_{\scriptscriptstyle C}, \; H \in \mathcal{F}(T^*Q)\\
\vspace{0.1cm}\displaystyle h(\Psi)= \int_{T^*Q}\bar\Psi \widehat{\cal L}_H\Psi\,\Lambda\\
{\rm i}\hbar \partial _t \Psi = \widehat{\cal L}_H\Psi
\end{array}
} \ar[ddrrr]|{\begin{array}{c}\text{Momentum map \eqref{Joli}}\\
\text{ for the group}\\
\operatorname{Diff}(T^*Q) \,\circledS\, \mathcal{F}(T^*Q, S^1)\\
\end{array}}
\ar[uurrr]|{\begin{array}{c}\text{Momentum map \eqref{KvHmomap}}\\
\text{for the subgroup}\\
\widehat{\operatorname{Diff}}_\omega(T^*Q)\\
\end{array}}
\ar[dd]|{\begin{array}{l}\Psi= Re^{{\rm i} S/\hbar}\end{array}} & & & \\
& & & & &\\
&*+[F-:<3pt>]{\begin{array}{l}\text{KvH-Madelung}\\
\text{equations \eqref{KvHMadelung1}}\\
\text{for $(S,R)$}
\end{array}}\ar[rrr] & & &
*+[F-:<3pt>]{
\begin{array}{l}
\vspace{0.1cm}\text{Madelung variables}\\
\vspace{0.1cm} ( \sigma , D)   \in \big( \mathfrak{X}(T^*Q) \,\circledS\, \mathcal{F}(T^*Q) \big) ^*
\end{array}
} \ar[uuuu]|{\begin{array}{c}\vspace{0.1cm}\text{Dual to the Lie algebra inclusion \eqref{LAincl}} \\
\iota:  \mathcal{F}(T^*Q)\hookrightarrow \mathfrak{X}(T^*Q) \,\circledS\, \mathcal{F}(T^*Q)\end{array}} &
}
\end{xy}
\end{framed}\vspace{-.5cm}\it
\caption{Schematic description of the various quantities appearing in Koopman-van Hove classical mechanics and some of the mappings between them.}
\label{figure1}
\end{figure}

\medskip\noindent
In more generality, we have seen how several quantities appearing in KvH classical mechanics are all interconnected between them and their relations are most often given by specific momentum maps associated to  particular diffeomorphisms of the prequantum bundle. The general picture of the variables appearing in KvH classical mechanics is presented in Figure \ref{figure1}, which displays also the role of the  various momentum maps presented so far.

\section{Hybrid quantum--classical dynamics}

\subsection{Quantum--classical wave equation\label{sec:CQeq}}
As mentioned earlier, the KvH framework leads naturally to the hybrid description of a coupled quantum--classical system. Indeed, one may start with the KvH equation for two particles and then apply geometric quantization to quantize one of them. 
Here, instead of quantizing observables, we follow an alternative procedure. As outlined by Klein \cite{Klein} in the case of one particle, this method transforms the KvH equation \eqref{KvH_eq} into the Schr\"odinger equation and here we restrict to consider Hamiltonians for the type $H=T+V$ (i.e. given by the sum of kinetic and potential energy). In one dimension Klein's method proceeds as follows: (1) write the one-particle KvH equation for $\Psi(x,\nu)$ with $H=m^{-1}\nu^2/2+V(x)$ and ${\mathcal{A}}=\nu\de x$, (2) restrict to consider solutions $\partial_\nu\Psi=0$, and (3) replace $\nu\to - {\rm i}\hbar\partial_x$. A direct verification shows that this yields the standard Schr\"odinger equation ${\rm i}\hbar\partial_t\Psi(x)=-({m^{-1}\hbar^2}/{2})\Delta\Psi  +V\Psi$. The condition $\partial_\nu\Psi=0$ corresponds to fixing a \emph{polarization} in geometric quantization \cite{Kirillov}, while the replacement $\nu\to-{\rm i}\hbar\partial_x$ corresponds to the usual canonical quantization prescription. 

At this point, a hybrid theory can be obtained by starting with the KvH equation for two particles and then applying Klein's method to quantize one of the particles.
This is precisely the approach adopted in \cite{BoGBTr}, which led to the following \emph{quantum--classical wave equation}
\begin{equation}\label{hybrid_KvH}
{\rm i}\hbar\partial_t\Upsilon=\{{\rm i}\hbar \widehat{H},\Upsilon\} + \big(\widehat{H} - {\mathcal{A}}\cdot X_{\widehat{H}}\big)\Upsilon
\,.
\end{equation}
Similar equations already appeared in \cite{boucher} and were rejected by the authors. Here, the phase-space function $\widehat{H}(z)$   takes values in the space of unbounded Hermitian operators on the quantum Hilbert space $\mathscr{H}_{\scriptscriptstyle Q}:=L^2( M)$. In addition, $\Upsilon \in L^2(T^*Q\times M)$ is a hybrid wavefunction depending on both the classical and the quantum coordinates, denoted by $z\in T^*Q$ and $x\in M$ respectively. Here, we assume that $M$ is endowed with a volume form $\mu$ so that the inner product and symplectic form on $L^2(T^*Q\times M)$ are defined by the immediate generalization of the classical definitions \eqref{inner_symplectic}. For convenience, here we shall denote the \emph{hybrid quantum--classical Hilbert space} by
\begin{equation}
\mathscr{H}_{\scriptscriptstyle QC}:=L^2(T^*Q\times M)
\,.
\end{equation}

It is useful to recall that the identification $\Upsilon (z,x)\simeq (\Upsilon(z))(x)$ yields the isometric isomorphism 
\begin{equation}\label{isom1}
\mathscr{H}_{\scriptscriptstyle QC}\simeq L^2(T^*Q; \mathscr{H}_{\scriptscriptstyle Q})
\,,
\end{equation}
where $L^2(T^*Q; \mathscr{H}_{\scriptscriptstyle Q})=\mathscr{H}_{\scriptscriptstyle C}\otimes \mathscr{H}_{\scriptscriptstyle Q}$ is the Bochner-Lebesgue space of $L^2$ functions on $T^*Q$ taking values in the quantum Hilbert space $\mathscr{H}_{\scriptscriptstyle Q}$; see, e.g., \cite[\S1.2]{HyNeVeWe2016}. Notice that the same approach yields the alternative isomorphism 
\beq
\mathscr{H}_{\scriptscriptstyle QC}\simeq L^2(M; \mathscr{H}_{\scriptscriptstyle C})
= \mathscr{H}_{\scriptscriptstyle C}\otimes \mathscr{H}_{\scriptscriptstyle Q}.
\label{isom2}
\eeq
Both isomorphisms \eqref{isom1} and \eqref{isom2} will be useful to compute integrals of the form{
\begin{align}
\int_M \bar{\Upsilon}(z',x){\Upsilon}(z,x)\,\mu:=&\ 
\int_M (\overline{\Upsilon(z')})(x)(\Upsilon(z))(x)\,\mu
\\
\int_{T^*Q} \bar{\Upsilon}(z,x'){\Upsilon}(z,x)\,\Lambda
:=&\ 
\int_{T^*Q} (\overline{\Upsilon(x')})(z)(\Upsilon(x))(z)\,\Lambda
\label{QDM}
\,.
\end{align}
In addition, these isomorphisms lead to defining the quantum adjoint $\Upsilon^\dagger(z)$ as follows. If we evaluate the square-integrable function $\Upsilon\in \mathscr{H}_{\scriptscriptstyle QC}$ at a fixed point $z\in T^*Q$, the isomorphism \eqref{isom1} yields another square-integrable function $\Upsilon(z) \in \mathscr{H}_{\scriptscriptstyle Q}$ in the quantum Hilbert space so that for each fixed $z$ the standard inner product $\langle\ | \ \rangle$ on $\mathscr{H}_{\scriptscriptstyle Q}$ induces a linear form $\Upsilon^\dagger(z)$ on $\mathscr{H}_{\scriptscriptstyle Q}$ given by
\begin{equation}\label{gigi}
\Upsilon^\dagger(z)\psi:=\langle\Upsilon(z)|\psi\rangle= \int_M \bar{\Upsilon}(z,x)\psi(x)\,\mu
\,,
\end{equation}
with $\psi\in \mathscr{H}_{\scriptscriptstyle Q}$.
For example, one writes $\Upsilon^\dagger(z)\Upsilon(z)=\|\Upsilon(z)\|^2$, where the norm $\|\ \|$ is induced by the inner product  $\langle\ | \ \rangle$ on $\mathscr{H}_{\scriptscriptstyle Q}$.
We shall call $\Upsilon^\dagger(z)$ the \emph{quantum adjoint of $\Upsilon$}, while an analogous procedure can evidently be used to define the classical variant.
}

 By construction, the \textit{hybrid Liouvillian operator}
\beq\label{hybliouv}
\widehat{\cal L}_{\widehat{H}}=\{{\rm i}\hbar \widehat{H},\ \} + \big(\widehat{H} - {\mathcal{A}}\!\cdot\! X_{\widehat{H}}\big)
\eeq
 is an unbounded Hermitian operator on $\mathscr{H}_{\scriptscriptstyle QC}$ so that the the  quantum--classical wave equation \eqref{hybrid_KvH}  reads ${\rm i}\hbar\partial_t\Upsilon=\widehat{\cal L}_{\widehat{H}}\Upsilon$ and the 
 hybrid wavefunction $\Upsilon$ undergoes unitary dynamics. 
Using local coordinates on $T^*Q$, the operator \eqref{hybliouv} is written as
\[
\widehat{\cal L}_{\widehat{H}}\Upsilon= {\rm i}\hbar  \big(\partial_{q^i}\widehat{H}\partial_{p_i}\Upsilon-\partial_{p_i}\widehat{H}\partial_{q^i}\Upsilon \big)+ \big(\widehat{H}- p_i \partial_{p_i}\widehat{H}\big)\Upsilon.
\]
Similarly to the injective correspondence $H\mapsto \widehat{\cal L}_{{H}}$ underlying the covariant Liouvillian operator \eqref{preqop}, the correspondence $\widehat{H} \rightarrow \widehat{\cal L}_{\widehat{H}}$ is also injective. Importantly, upon considering the immediate generalization of the symplectic form in \eqref{inner_symplectic}, we notice that the quantum--classical wave equation \eqref{hybrid_KvH} is  Hamiltonian with the following Hamiltonian functional expressed in terms of the quantum Liouvillian:
\begin{equation}\label{hybHam}
h(\Upsilon)=\int_{T^*Q}\!\big\langle\Upsilon\big|\widehat{\cal L}_{\widehat{H}}\Upsilon\big\rangle\,\Lambda
=
\int_{T^*Q}\!\int_M\big(\bar \Upsilon\,\widehat{\cal L}_{\widehat{H}}\,\Upsilon\big)\,\Lambda\wedge \mu
\,.
\end{equation}
While hybrid Liouvillian operators do not comprise a Lie algebra structure, the next section  presents their general algebraic properties.

\subsection{Algebra of hybrid Liouvillian operators\label{sec:algebra}}

While the covariant Liouvillian operators defined in \eqref{preqop} possess a Lie algebra structure given by $[\widehat{\cal L}_F,\widehat{\cal L}_K]={\rm i}\hbar\widehat{\cal L}_{\{F,K\}}$, no such structure is available for the hybrid Liouvillian operators. The latter satisfy obvious identities that can be written upon introducing the convenient notation $A_C\in \mathcal{F}(T^*Q)$ for classical observables and $\widehat{A}_Q\in \operatorname{Her}(\mathscr{H}_{\scriptscriptstyle Q})$ for quantum observables, while $\widehat{A}\in\mathcal{F}(T^*Q,\operatorname{Her}(\mathscr{H}_{\scriptscriptstyle Q}))$ denotes a hybrid observable. Here, $\operatorname{Her}(\mathscr{H}_{\scriptscriptstyle Q})$ denotes the space of Hermitian operators on  $\mathscr{H}_{\scriptscriptstyle Q}$, so that $\mathcal{F}(T^*Q,\operatorname{Her}(\mathscr{H}_{\scriptscriptstyle Q}))$ is the space
of phase-space functions taking values in the space  $\operatorname{Her}(\mathscr{H}_{\scriptscriptstyle Q})$ of quantum observables.
For example, with this notation we have  ${\widehat{\cal L}}_{\widehat{A}_{Q}}=\widehat{A}_{Q}$. More generally,  one has the obvious identities
\[
{\widehat{\cal L}}_{{\widehat{A}_{Q}}{\widehat{B}}}={\widehat{A}_Q}{\widehat{\cal L}}_{\widehat{B}}
\,,\qquad\qquad
\big[{\widehat{\cal L}}_{\widehat{A}_Q},{\widehat{\cal L}}_{\widehat{B}}\big]={\widehat{\cal L}}_{[{\widehat{A}_Q},{\widehat{B}}]}
\,,\qquad\qquad
\big[{\widehat{\cal L}}_{{A}_C},{\widehat{\cal L}}_{\widehat{B}}\big]={\rm i}\hbar{\widehat{\cal L}}_{\{{{A}_C},{\widehat{B}}\}}
\,,
\]
as well as
\begin{align*}
\big[{\widehat{\cal L}}_{\widehat{A}_Q{A}_C},{\widehat{\cal L}}_{\widehat{B}_Q}\big]
=
&\ 
{\widehat{\cal L}}_{\,[{\widehat{A}_Q},{\widehat{B}_Q}]{A}_C}
\,,\qquad\qquad
\big[{\widehat{\cal L}}_{\widehat{A}_Q{A}_C},{\widehat{\cal L}}_{{B}_C}\big]={\rm i}\hbar{\widehat{\cal L}}_{\{A_C,{B}_C\}\widehat{A}_Q}
\,.
\end{align*}

In addition to the above algebraic rules,  here it may be useful to report a further remarkable relation. To this end, we start by introducing the conjugate of an operator as follows. 

\medskip

\begin{definition}[Conjugate operator]
Let $\mathscr{H}$ a complex Hilbert space and let $\widehat{A}:\mathscr{H}\to\mathscr{H}$ be a linear operator. Then, the conjugate operator $\bar{A}$ of $\widehat{A}$ is defined by the relation $\bar{A}u:= \overline{\widehat{A} \bar{u}}$ for any $u\in \mathscr{H}$. 
\end{definition}
\noindent Here, no confusion should arise from  adopting the notation $\bar{A}$ in place of $\,\overline{\!\widehat{A}\,}\!$\,. For example, if  $\mathscr{H}=\mathscr{H}_{\scriptscriptstyle Q}$ and $\widehat{A}$ is an integral operator with kernel ${\cal K}_{\widehat{A}}(x,x')$ (or `matrix element', in  physics terminology), then we have ${\cal K}_{\bar{A}}(x,x')=\overline{{\cal K}_{\widehat{A}}(x,x')}$. 
Also, in analogy with the definition of the transpose of a linear mapping, recalling that ${\mathscr{H}_{\scriptscriptstyle QC}=\mathscr{H}_{\scriptscriptstyle C}\otimes\mathscr{H}_{\scriptscriptstyle Q}}$ is a tensor product space, we define the \emph{quantum transpose} of a linear operator ${\widehat{\cal L}: \mathscr{H}_{\scriptscriptstyle QC} \rightarrow \mathscr{H}_{\scriptscriptstyle QC}}$  as the partial transpose with respect to the factor $\mathscr{H}_{\scriptscriptstyle C}$. The intrinsic definition is given by

\begin{definition}[Quantum transpose]
Let ${\widehat{\cal L}: \mathscr{H}_{\scriptscriptstyle QC} \rightarrow \mathscr{H}_{\scriptscriptstyle QC}}$ be a linear operator on the tensor-product Hilbert space ${\mathscr{H}_{\scriptscriptstyle QC}=\mathscr{H}_{\scriptscriptstyle C}\otimes\mathscr{H}_{\scriptscriptstyle Q}}$. Then, the quantum transpose $\widehat{\mathcal{L}}^{\,\mathsf{T}}$ of  $\widehat{\cal L}$ is the operator
\begin{equation}\label{quantum_T}
\big\langle \widehat{\mathcal{L}}^{\,\mathsf{T}}(\Psi_2\psi_2)\,|\, \Psi_1 \psi_1 \big\rangle := \left\langle {\bar\Psi}_1 \psi_2\,|\, \bar{\cal L}(\bar{\Psi}_2 \psi_1) \right\rangle ,
\end{equation}
for all $\Psi_1,\Psi_2\in \mathscr{H}_{\scriptscriptstyle C}$ and $\psi_1,\psi_2\in \mathscr{H}_{\scriptscriptstyle Q}$. 
\end{definition}
Notice the position of the indices $1$ and $2$ in this definition. In the case of an integral operator $\widehat{\mathcal{L}}$ with kernel ${\cal K}_{ \widehat{\mathcal{L}}}(z,z',x,x')$, we have ${\cal K}_{ \widehat{\mathcal{L}}^{\,\mathsf{T}}}(z,z',x,x')={\cal K}_{ \widehat{\mathcal{L}}}(z,z',x',x)$.
At this point, with the definitions above, we have the following result:
\begin{theorem}[Remarkable algebraic relation]
 Let $\widehat{A}(z)$ and $\widehat{B}(z)$ be  two hybrid observables in the space $\mathcal{F}(T^*Q,\operatorname{Her}(\mathscr{H}_{\scriptscriptstyle Q}))$ of phase-space functions taking values in the space of Hermitian operators $\operatorname{Her}(\mathscr{H}_{\scriptscriptstyle Q})$ on the quantum Hilbert space $\mathscr{H}_{\scriptscriptstyle Q}$. Then, their associated Liouvillian operators ${\widehat{\cal L}}_{\widehat{A}}$ and ${\widehat{\cal L}}_{\widehat{B}}$ satisfy the following identity:
\beq\label{RR}
\big[{\widehat{\cal L}}_{\widehat{A}},{\widehat{\cal L}}_{\widehat{B}}\big]+
\big[{\widehat{\cal L}}_{\bar{A}},{\widehat{\cal L}}_{\bar{B}}\big]^{\!{\sf T}}
=\ {\rm i}\hbar
{\widehat{\cal L}}_{\{{\widehat{A}},{\widehat{B}}\}-\{{\widehat{B}},{\widehat{A}}\}}.
\eeq
\end{theorem}
\paragraph{Proof.}
It will be convenient to distinguish between the the inner products $\langle \ | \ \rangle_{\scriptscriptstyle C}$, $\langle \ | \ \rangle_{\scriptscriptstyle Q}$, and $\langle \ | \ \rangle_{\scriptscriptstyle QC}$ on the different Hilbert spaces $\mathscr{H}_{\scriptscriptstyle C}$, $\mathscr{H}_{\scriptscriptstyle Q}$, and $\mathscr{H}_{\scriptscriptstyle QC}$, respectively.
We note that for $\Theta, \Psi \in \mathscr{H}_{\scriptscriptstyle C}$ and $\theta, \psi \in \mathscr{H}_{\scriptscriptstyle Q}$, we have
\begin{equation}\label{formula_L_meanfields}
\Big\langle \Psi \psi \Big | \widehat{\cal L}_{\widehat A \,} (\Theta \theta)\Big\rangle_{ \scriptscriptstyle\! QC}= \Big\langle \Psi \Big | \widehat{\cal L}_{{\langle \psi |\widehat A \theta\rangle}_{\scriptscriptstyle Q\!}} \Theta\Big\rangle_{ \scriptscriptstyle\! C},
\end{equation}
where on the right hand side $\widehat{\cal L}_{{\langle \psi |\widehat A \theta\rangle}_{\scriptscriptstyle Q}}$ denotes the classical covariant Liouvillian operator \eqref{preqop} associated to the function  $z\in T^*Q \mapsto \langle \psi |\widehat H(z) \theta\rangle_{\scriptscriptstyle Q}\in \mathbb{C}$. In the remainder of this section, it will be convenient to use Dirac's notation for vectors in the quantum Hilbert space $\mathscr{H}_{\scriptscriptstyle Q}$; for example, we replace 
${\langle \psi |\widehat A \theta\rangle}_{\scriptscriptstyle Q\!}$ by ${\langle \psi |{\widehat A}| \theta\rangle}_{\scriptscriptstyle Q\!}$ in \eqref{formula_L_meanfields}.
  Then,
choosing a sequence $|\alpha\rangle\in \mathscr{H}_{\scriptscriptstyle Q}$ such that $\int |\alpha\rangle \langle\alpha|\, {\rm d}\alpha={\rm id}_{ \mathscr{H}_{\scriptscriptstyle Q}}$, we can write
\begin{align*}
\Big\langle \Psi \psi \Big | \big[\widehat{\cal L}_{\widehat{A}},\widehat{\cal L}_{\widehat{B}}\big] (\Theta \theta)\Big\rangle_{\scriptscriptstyle\! QC}
=&
 \int\! \Big\langle \Psi \Big | \left(\widehat{\cal L}_{{\langle \psi |{\widehat A}| \alpha\rangle}_{\scriptscriptstyle Q}}\widehat{\cal L}_{{\langle \alpha |{\widehat B}| \theta\rangle}_{\scriptscriptstyle Q}\!}- \widehat{\cal L}_{{\langle \psi |{\widehat B}| \alpha\rangle}_{\scriptscriptstyle Q}}\widehat{\cal L}_{{\langle \alpha |{\widehat A}| \theta\rangle}_{\scriptscriptstyle Q}} \right)\Theta\Big\rangle_{\scriptscriptstyle C\,} {\rm d}\alpha
\\
=&\int\!  \Big\langle \Psi \Big | \left( \widehat{\cal L}_{{\langle \psi |{\widehat A}| \alpha\rangle}_{\scriptscriptstyle Q}}\widehat{\cal L}_{{\langle \alpha |{\widehat B}| \theta\rangle}_{\scriptscriptstyle Q}\!} -  \widehat{\cal L}_{{\langle \alpha |{\widehat B}| \theta\rangle}_{\scriptscriptstyle Q}}\widehat{\cal L}_{{\langle \psi |{\widehat A}| \alpha\rangle}_{\scriptscriptstyle Q}\!} \right)\Theta \Big\rangle_{\scriptscriptstyle \! C\,}{\rm d}\alpha
\\
&+\int\!  \Big\langle \Psi \Big | \left(   \widehat{\cal L}_{{\langle \alpha |{\widehat B}| \theta\rangle}_{\scriptscriptstyle Q}}\widehat{\cal L}_{{\langle \psi |{\widehat A}| \alpha\rangle}_{\scriptscriptstyle Q}}-\widehat{\cal L}_{{\langle \psi |{\widehat B}| \alpha\rangle}_{\scriptscriptstyle Q}}\widehat{\cal L}_{{\langle \alpha |{\widehat A}| \theta\rangle}_{\scriptscriptstyle Q}\!}\right)\Theta \Big\rangle_{\scriptscriptstyle \! C\,}{\rm d}\alpha
\\
& +\int\!  \Big\langle \Psi \Big | \left(   \widehat{\cal L}_{{\langle \alpha |{\widehat A}| \theta\rangle}_{\scriptscriptstyle Q}} \widehat{\cal L}_{{\langle \psi |{\widehat B}| \alpha\rangle}_{\scriptscriptstyle Q}\!} -   \widehat{\cal L}_{{\langle \alpha |{\widehat A}| \theta\rangle}_{\scriptscriptstyle Q}} \widehat{\cal L}_{{\langle \psi |{\widehat B}| \alpha\rangle}_{\scriptscriptstyle Q}} \right)\Theta \Big\rangle_{ \scriptscriptstyle\! C\,}{\rm d}\alpha
\\
=&\int\!  \Big\langle \Psi \Big | {\rm i}\hbar  \widehat{\cal L}_{\left \{ {\langle \psi |{\widehat A}| \alpha\rangle}_{\scriptscriptstyle Q}, \,{\langle \alpha |{\widehat B}| \theta\rangle}_{\scriptscriptstyle Q}\right\}}\Theta -   {\rm i}\hbar  \widehat{\cal L}_{\left \{ {\langle \psi |{\widehat B}| \alpha\rangle}_{\scriptscriptstyle Q},\, {\langle \alpha |{\widehat A}| \theta\rangle}_{\scriptscriptstyle Q}\right\}}\Theta\Big\rangle_{\scriptscriptstyle\!  C\,}{\rm d}\alpha
\\
&+ \int\!  \Big\langle \Psi \Big |  \widehat{\cal L}_{{\langle \alpha |{\widehat B}| \theta\rangle}_{\scriptscriptstyle Q}}\widehat{\cal L}_{{\langle \psi |{\widehat A}| \alpha\rangle}_{\scriptscriptstyle Q}} \Theta- \widehat{\cal L}_{{\langle \alpha |{\widehat A}| \theta\rangle}_{\scriptscriptstyle Q}} \widehat{\cal L}_{{\langle \psi |{\widehat B}| \alpha\rangle}_{\scriptscriptstyle Q}} \Theta \Big\rangle_{ \scriptscriptstyle\! C\,}{\rm d}\alpha,
\end{align*}
where in the third equality we used $[\widehat{\cal L}_H,\widehat{\cal L}_F]={\rm i}\hbar\widehat{\cal L}_{\{H,F\}}$ for $\widehat{\cal L}$ given in \eqref{preqop}. The first two terms in the last equality can be written as
\[
\Big\langle \Psi \Big | {\rm i}\hbar  \widehat{\cal L}_{ {\langle \psi |  \{\widehat A ,\widehat B \}-  \{\widehat B ,\widehat A \}| \theta\rangle}_{\scriptscriptstyle Q}}\Theta \Big\rangle_{\scriptscriptstyle\! C}=\Big\langle \Psi \psi \Big | {\rm i}\hbar  \widehat{\cal L}_{  \{\widehat A ,\widehat B \}-  \{\widehat B ,\widehat A \}} \Theta \theta \Big\rangle_{\scriptscriptstyle\! QC}
\]
while the last two terms are
\begin{align*}
&\int \! \Big\langle \Psi \Big |  \widehat{\cal L}_{{\langle \bar\theta |{\bar B}| \bar\alpha\rangle}_{\scriptscriptstyle Q}}\widehat{\cal L}_{{\langle \bar \alpha |{\bar A} |\bar\psi\rangle}_{\scriptscriptstyle Q}} \Theta- \widehat{\cal L}_{{\langle \bar\theta |{\bar A}| \bar\alpha\rangle}_{\scriptscriptstyle Q}} \widehat{\cal L}_{{\langle  \bar\alpha |{\bar B}| \bar\psi\rangle}_{\scriptscriptstyle Q}} \Theta \Big\rangle_{\scriptscriptstyle\! C\,}{\rm d}\alpha 
= \ 
\Big\langle \Psi \bar\theta\, \Big | \big[\widehat{\cal L}_{\bar B}, \widehat{\cal L}_{\bar A}\big](\Theta\bar\psi) \Big\rangle_{\scriptscriptstyle\! QC}\\
&\qquad\qquad=\Big\langle  \,\overline{\!\big[\widehat{\cal L}_{\bar B}, \widehat{\cal L}_{\bar A}\big]\!}\,(\bar \Theta\psi)   \Big | \bar\Psi \theta \Big\rangle_{\scriptscriptstyle\! QC} 
= \  
\Big\langle \Psi \psi \Big |  \big[\widehat{\cal L}_{\bar B}, \widehat{\cal L}_{\bar A}\big]^{\!\mathsf{T}}\Theta \theta  \Big\rangle_{ \scriptscriptstyle\!QC}
\end{align*}
by using \eqref{quantum_T}. These relations are satisfied for all $\Psi,\Theta\in \mathscr{H}_{\scriptscriptstyle C}$ and all $\psi, \theta\in\mathscr{H}_{\scriptscriptstyle Q}$. In particular they hold for any orthonormal bases $(\Psi_i)_{i\in I}$ and $(\psi_j)_{j\in J}$ of $\mathscr{H}_{\scriptscriptstyle C}$ and $\mathscr{H}_{\scriptscriptstyle Q}$, respectively. Since $(\Psi_i\otimes \psi_j)_{(i,j)\in I \times J}$ is an orthonormal basis of ${\mathscr{H}_{\scriptscriptstyle QC}=\mathscr{H}_{\scriptscriptstyle C}\otimes\mathscr{H}_{\scriptscriptstyle Q}}$, we obtain \eqref{RR}. 
\qquad\qquad$\blacksquare$

\medskip

We conclude this section by presenting the equivariance properties of hybrid Liouvillians.  These properties will be used in the next section to obtain a physically  more relevant result.
\begin{lemma}[Equivariance]\label{lemmaequiv} Let $\widehat{A}(z)$  be  a hybrid observable in the space $\mathcal{F}(T^*Q,\operatorname{Her}(\mathscr{H}_{\scriptscriptstyle Q}))$ of phase-space functions taking values in the space of Hermitian operators $\operatorname{Her}(\mathscr{H}_{\scriptscriptstyle Q})$ on the quantum Hilbert space $\mathscr{H}_{\scriptscriptstyle Q}$. Also, let ${\mathcal{U}}(\mathscr{H}_{\scriptscriptstyle Q})$ denote the group of unitary operators on  $\mathscr{H}_{\scriptscriptstyle Q}$ and $U_{(\eta,e^{{\rm i}\theta})}$ denote the van Hove unitary operator  \eqref{cl_prop}. Then, the   Liouvillian  $\widehat{\cal L}_{\widehat{A}}$ associated to $\widehat{A}(z)$ satisfies
\beq\label{ignazio}
U_{(\eta,e^{{\rm i}\theta})}^\dagger\widehat{\cal L}_{\widehat{A}}U_{(\eta,e^{{\rm i}\theta})}=\widehat{\cal L}_{\eta^*\!\widehat{A}}
\,,\qquad\forall\; (\eta,e^{{\rm i}\theta})\in \widehat{\operatorname{Diff}}_\omega(T^*Q)
\,;
\eeq 
\beq\label{cov2L}
U^\dagger\widehat{\cal L}_{\widehat{A}}U=\widehat{\cal L}_{U^\dagger\widehat{A}U}
\,,\qquad\forall\;\widehat{{U}}\in {\mathcal{U}}(\mathscr{H}_{\scriptscriptstyle Q})
\,.
\eeq
\end{lemma}
\paragraph{Proof.} The  relation \eqref{ignazio} is proved in Appendix \ref{app:equiv}, while \eqref{cov2L} follows by immediate verification.\qquad\qquad$\blacksquare$

\medskip
\noindent
On one hand, the relation \eqref{ignazio} is a direct extension of \eqref{equivariance_L_H} to the case of hybrid Liouvillians. This is a natural consequence of the fact that the van Hove representation $U_{(\eta, e^{{\rm i}\theta})}$ in \eqref{cl_prop} does not involve quantum degrees of freedom.
On the other hand, one also has equivariance under unitary transformations of the quantum Hilbert space space.
In the next section, these equivariance relations will be shown to apply also to  a hybrid density operator extending the classical Liouville density as well as the von Neumann's celebrated density matrix in quantum theory.

\subsection{The hybrid density operator\label{sec:hybden}}

As shown in \cite{BoGBTr}, the Hamiltonian structure of the quantum--classical wave equation \eqref{hybrid_KvH} leads to defining a hybrid density operator for the computation of expectation values. Indeed, the latter can be identified by rewriting the Hamiltonian functional \eqref{hybHam} by using integration by parts as follows:
\beq\label{Dintro}
h(\Upsilon)=\int_{T^*Q}\!\big\langle\Upsilon\big|\widehat{\cal L}_{\widehat{H}}\Upsilon\big\rangle\,\Lambda
=\operatorname{Tr}\int _{T^*Q\!}\widehat{H}(z)\widehat{\cal D}(z)\,\Lambda
\,.
\eeq
Here, in analogy to the expression \eqref{KvHmomap} of the classical Liouville density, the   hybrid density operator $\widehat{\cal D}$ is given as
\begin{equation}\label{hybridDenOp}
\widehat{\cal D}=\Upsilon\Upsilon^\dagger - \operatorname{div}\!\big( \mathbb{J}{\mathcal{A}} \Upsilon \Upsilon^\dagger\big) + {\rm i}\hbar\{\Upsilon,\Upsilon^\dagger\}\,,
\end{equation}
so that $\operatorname{Tr}\int_{\scriptscriptstyle T^*\!Q}\widehat{\cal D}\,\Lambda=1$. Again, here the divergence is taken relative to the Liouville volume form $\Lambda$ on $T^*Q$. The hybrid density operator is defined in such a way that its application to a quantum wavefunction $\psi\in \mathscr{H}_{\scriptscriptstyle Q}$ reads
\begin{equation}\label{hatD_psi}
\widehat{\cal D}(z)\psi= \Upsilon(z) \langle\Upsilon(z)|\psi\rangle + \partial_{p_i} (p_i \langle\Upsilon(z)|\psi\rangle\Upsilon (z) )+{\rm i}\hbar \{\Upsilon(z) , \langle\Upsilon(z)|\psi\rangle\}
\,,
\end{equation}
where we recall \eqref{gigi}.
As usual, by appropriately restricting the space of wavefunctions $\Upsilon$, the associated hybrid density operator $\widehat{\cal D}$ is an operator-valued density on $T^*Q$ taking values in the space of trace-class Hermitian operators on $L^2(M)$. We write $\widehat{\cal D}\in\operatorname{Den}(T^*Q, \operatorname{Her}(\mathscr{H}_{\scriptscriptstyle Q}))$. In particular, $\widehat{\cal D}$ is an integral operator with kernel (or `matrix element')
\begin{align}\nonumber
{\cal K}_{\widehat{\cal D}}(z;x,x')=&\, \Upsilon(z,x)\bar{\Upsilon}(z,x')- \partial_{p_i} \big(p_i \Upsilon(z,x)\bar{\Upsilon}(z,x')\big)
\\
&\,
+{\rm i}\hbar \big( \partial_{q^i}\Upsilon(z,x)\partial_{p_i}\bar{\Upsilon}(z,x') - \partial_{p_i}\Upsilon(z,x)\partial_{q^i}\bar{\Upsilon}(z,x')\big).
\label{kappa_D}
\end{align}

Given the hybrid density operator $\widehat{\cal D}$, one computes the quantum density operator 
\beq\label{QuantDensMat}
\hat\rho:=\int_{\scriptscriptstyle T^*\!Q\!}\widehat{\cal D}(z)\,\Lambda = \int_{T^*Q} \Upsilon(z)\Upsilon^\dagger(z)\,\Lambda
\,,
\eeq
so that the quantum probability density in configuration space is obtained as
\begin{equation}\label{rhoq}
\rho_q(x)=\int_{T^*Q}{\cal K}_{\widehat{\mathcal{D}}}(z;x,x) \Lambda
=
\int_{T^*Q} |\Upsilon(z,x)|^2\,\Lambda\,.
\end{equation}
On the other hand,  the classical density reads
$
\rho_c(z)=\operatorname{Tr}\widehat{\cal D}(z)=\int_{\scriptscriptstyle M}{\cal K}_{\widehat{\mathcal{D}}}(z;x,x) \mu$.
Here the trace is  computed only with respect to the quantum degrees of freedom so that, \makebox{upon using \eqref{kappa_D}},
\begin{align}\label{rhoc}
\rho_c(z)=&
\operatorname{Tr}\widehat{\cal D}(z) =\int_M  \Big[|\Upsilon(z,x)|^2-  \partial_{p_i\!} \big(p_i|\Upsilon(z,x)|^2\big) + {\rm i}\hbar \{\Upsilon, \bar\Upsilon\}(z,x)\Big]\,\mu.
\end{align}

We now move on to discuss expectation values. It is evident that the second equality in \eqref{Dintro} holds upon replacing $\widehat{H}(z)$  by any hybrid quantum--classical observable $\widehat{A}(z)$, whose expectation value $\langle\widehat{A}\rangle$ can then be written as 
\[
\langle\widehat{A}\rangle=\operatorname{Tr}\int_{T^*Q}\widehat{A}(z)\widehat{\cal D}(z)\,\Lambda=\int_{T^*Q}\!\big\langle\Upsilon\big|\widehat{\cal L}_{\widehat{A}}\Upsilon\big\rangle\,\Lambda
\,.
\]
This relation extends  the classical case \eqref{expvalc}. Again, we notice the difference from the relations appearing in the purely quantum formalism. The usual quantum expectation value is however recovered naturally in the purely quantum case, since the relation $\widehat{\cal L}_{\widehat{A}_Q}=\widehat{A}_Q$ from Section \ref{sec:algebra} for a  quantum observable $\widehat{A}_Q\in \operatorname{Her}(\mathscr{H}_{\scriptscriptstyle Q})$ implies $\langle\widehat{A}_Q\rangle=\int_{\scriptscriptstyle T^*\!Q}\langle\Upsilon|\widehat{A}_Q\Upsilon\rangle\,\Lambda$. 


\subsection{Equivariance of the hybrid density operator\label{sec:equiv}}
Another feature of the hybrid density operator $\widehat{\cal D}$ is the equivariance property of its defining mapping ${\Upsilon\mapsto\widehat{\cal D}(\Upsilon)}$ in \eqref{hybridDenOp} under both quantum and classical transformations. 
More specifically, we have the following result.

\begin{theorem}[Equivariance of the hybrid density operator] \quad\!\!\!\!\!   Let { ${\mathcal{U}}(\mathscr{H}_{\scriptscriptstyle Q})$ denote the} \makebox{group of unitary operators} on  $\mathscr{H}_{\scriptscriptstyle Q}$ and $U_{(\eta,e^{{\rm i}\theta})}$ denote the  unitary operator \eqref{cl_prop} corresponding to the van Hove representation of $(\eta,e^{{\rm i}\theta})\in\widehat{\operatorname{Diff}}_\omega(T^*Q)$ on $\mathscr{H}_{\scriptscriptstyle QC}$. Also, let the hybrid density operator $\widehat{\cal D}\in ({\cal F}(T^*Q,\operatorname{Her}(\mathscr{H}_{\scriptscriptstyle Q})))^*$ be defined as in \eqref{hatD_psi}.  Then, one has
\beq
\widehat{\cal D}(U_{(\eta,e^{{\rm i}\theta})}\Upsilon)= \eta_*\big(\widehat{\cal D}(\Upsilon)\big)
\,,\qquad\forall\; (\eta,e^{{\rm i}\theta})\in \widehat{\operatorname{Diff}}_\omega(T^*Q)
\,.
\label{cov1}
\eeq
\beq\label{cov2}
\widehat{\cal D}(\widehat{{U}}\Upsilon)= \widehat{{U}}\widehat{\cal D}(\Upsilon)\widehat{{U}}^\dagger
\,,\qquad\forall\;\widehat{{U}}\in {\mathcal{U}}(\mathscr{H}_{\scriptscriptstyle Q})
\,.
\eeq

\end{theorem}
\paragraph{Proof.} 
The  relation \eqref{cov1} can be verified by pairing $\widehat{\cal D}(U_{(\eta,e^{{\rm i}\theta})}\Upsilon)$ against a hybrid observable $\widehat{A}\in{\cal F}(T^*Q,\operatorname{Her}(\mathscr{H}_{\scriptscriptstyle Q}))$ as follows:
\begin{align*}
\operatorname{Tr}\int_{T^*Q}\widehat{A}\widehat{\cal D}(U_{(\eta,e^{{\rm i}\theta})}\Upsilon)\,\Lambda=&\int_{T^*Q}\!\big\langle U_{(\eta,e^{{\rm i}\theta})}\Upsilon\big|\widehat{\cal L}_{\widehat{A}}U_{(\eta,e^{{\rm i}\theta})}\Upsilon\big\rangle\,\Lambda
\\
=&\int_{T^*Q}\!\big\langle \Upsilon\big|U_{(\eta,e^{{\rm i}\theta})}^\dagger\widehat{\cal L}_{\widehat{A}}U_{(\eta,e^{{\rm i}\theta})}\Upsilon\big\rangle\,\Lambda
\\
=&\int_{T^*Q}\!\big\langle \Upsilon\big|\widehat{\cal L}_{\eta^*\!\widehat{A}}\Upsilon\big\rangle\,\Lambda
\\
=&\ 
\operatorname{Tr}\int_{T^*Q}\widehat{\cal D}(\Upsilon)\ \eta^*\!\widehat{A}\ \Lambda
\\
=&\ 
\operatorname{Tr}\int_{T^*Q} \widehat{A}\ \eta_*\big(\widehat{\cal D}(\Upsilon)\big)\ \Lambda
\,,
\end{align*}
where we used the relation \eqref{ignazio}. In addition,  \eqref{cov2} follows by construction from the definition \eqref{hybridDenOp}. \qquad\qquad$\blacksquare$

\medskip\noindent
The equivariance properties \eqref{cov1}-\eqref{cov2} of the hybrid density operator under both classical and quantum transformations have long been sought in the theory of hybrid quantum--classical systems \cite{boucher} and stand as one of the key geometric properties of the present construction. The equivariance properties \eqref{cov1}-\eqref{cov2} also reflect in the dynamics of both the classical distribution \eqref{rhoc} and  the quantum density matrix \eqref{QuantDensMat}, which read respectively \cite{BoGBTr}
\beq\label{joli}
\frac{\partial \rho_c}{\partial t}= \operatorname{Tr}\{\widehat{H},\widehat{\cal D}\}
\,,\qquad\qquad
i\hbar\frac{\partial \hat\rho}{\partial t}= \!\int_{T^*Q\,} [\widehat{H},\widehat{\cal D}]\ \Lambda
\eeq
As pointed out in \cite{BoGBTr}, pure state solutions are prevented by the density matrix evolution and this property is known as \emph{decoherence} in the physics terminology. In addition, classical point trajectories are also lost in the general case  of quantum--classical interaction, since the first equation in \eqref{joli} does not possess delta-like Klimontovich solutions (that is, classical pure states). While not completely surprising, the absence of classical particle trajectories in hybrid dynamics raises questions about the meaning of the word `classical' in this context. Classical motion is identified with a Hamiltonian flow producing characteristic curves representing particle trajectories. Then, the question emerges whether    any feature of a classical Hamiltonian flow can still be identified in hybrid dynamics. In this paper, we address this question by extending the Lagrangian trajectories from quantum hydrodynamics to hybrid quantum--classical systems. To this purpose, the following sections will apply the Madelung transform to equation \eqref{CQwaveq}. As a result, we shall present a hybrid generalization of Bohmian trajectories in terms of Lagrangian paths, which will be discussed in terms of their Hamiltonian structure and the corresponding momentum maps.

\section{Hybrid Madelung equations\label{sec:HybMad}}

In the remainder of this paper, we shall restrict to consider hybrid Hamiltonians of the type
\begin{equation}\label{QHam}
\widehat{H}(q,p, {x})=-\frac{\hbar^2}{2m}\Delta_x + \frac1{2M}|p|^2 + V(q,x)
\,,
\end{equation}
thereby ignoring the possible presence of magnetic fields. Here $\Delta_x$ is the Laplacian on $M$ associated to a given Riemannian metric and the norm $|p|$ is given with respect to a Riemannian metric on $Q$. In this case, the hybrid quantum--classical wave equation \eqref{hybrid_KvH} reads
\begin{equation}
{\rm i}\hbar\partial_t\Upsilon=-\left(L_I+\frac{\hbar^2}{2m}\Delta_x\right)\Upsilon+{\rm i}\hbar\left\{H_I,\Upsilon\right\}
\,,
\label{CQwaveq}
\end{equation}
where we have defined the following scalar functions $L_I, H_I$ on the hybrid space $T^*Q\times M$:
\beq\label{intHam}
H_I(q,p,{x}): = \frac1{2M}|p|^2 + V(q,x)
\,,\qquad\qquad
L_I(q,p,{x}): = \frac1{2M}|p|^2 - V(q,x)
\,.
\eeq
These are respectively the classical Hamiltonian and Lagrangian both augmented by the presence of the interaction potential. As we shall see, these quantities play a key role in the geometry of hybrid quantum--classical systems.

\subsection{Quantum--classical Madelung transform}

In this section we extend the usual Madelung transformation from quantum mechanics to the more general setting of coupled quantum--classical systems. The Madelung transform was already applied to KvH classical mechanics in Section \ref{sec:KvHMadelung}, while the equations \eqref{MadelungEqs2} for standard quantum hydrodynamics appeared originally in Madelung's work \cite{Madelung}. We emphasize that here we focus on Madelung's original approach by invoking a single-valued phase function. The possibility of  multi-valued quantum phase functions leading to topological singularities  \cite{Dirac} and nontrivial holonomy was emphasized in \cite{Wallstrom}  and will not be considered in the present context. See also \cite{FoTr} for a geometric dynamical treatment of nontrivial holonomy in quantum hydrodynamics.

In order to apply the Madelung transform to the hybrid setting, we write the hybrid wavefunction in polar form, that is
\begin{equation}\label{polarform}
\Upsilon(t,z,x) = \R(t,z,x) e^{{\rm i}{\cal S}(t,z,x)/\hbar}
,
\end{equation}
where calligraphic fonts are used to distinguish the hybrid case from the previous purely classical case treated in Section \ref{sec:KvHMadelung}. Then, the quantum--classical wave equation \eqref{CQwaveq} produces the following dynamics for the hybrid amplitude and phase
\begin{align}\label{HybHJ}
&\frac{\partial  {\cal S}}{\partial t}+\frac{|\nabla_x {\cal S}|^2}{2m}-\frac{\hbar^2}{2m}\frac{\Delta_x  \R}{ \R}= L_I+\left\{H_I, {\cal S}\right\}
\\
&\frac{\partial  \R}{\partial t}+\frac{1}{2m\R}\operatorname{div}_x( \R^2\nabla_x {\cal S})= \left\{H_I,  \R\right\},\label{HybHJR}
\end{align}
where the operators $\nabla_x$, $\operatorname{div}_x$, and $\Delta_x= \operatorname{div}_x\nabla_x$ are defined in terms of the Riemannian metric on $M$. 
Each equation carries the usual quantum terms in \eqref{MadelungEqs}-\eqref{MadelungEqsbis} on the left-hand side, while the terms arising from KvH classical dynamics appear  on the right-hand side (see  equations \eqref{KvHMadelung1}-\eqref{KvHMadelung2}). In the absence of classical degrees of freedom, the first equation simply recovers the so called \emph{quantum Hamilton-Jacobi equation} \eqref{MadelungEqs}. 

We observe that \eqref{HybHJ} can be written in Lie derivative form as follows:
\begin{align}\label{HybMad1a}
&\left(\frac{\partial }{\partial t}+\pounds_\mathsf{X}\right) {\cal S} =\mathscr{L}
\,.
\end{align}
Here, $\mathsf{X}$ is the hybrid velocity vector field on $T^*Q\times M$ given by
\begin{equation}
\mathsf{X}(z,x)=\left(X_{H_I}(z,x),\frac{\nabla_x{\cal S}(z,x)}m\right)\,,
\label{HybridVecF}
\end{equation}
$X_{H_I}$ being the $x$-dependent Hamiltonian vector field associated to $H_I$ on $(T^*Q,\omega)$. Moreover, we have defined the (time-dependent) hybrid quantum--classical Lagrangian
\[
\mathscr{L}(t,z,x):=L_I + \frac{|\nabla_x {\cal S}|^2}{2m} + \frac{\hbar^2}{2m}\frac{\Delta_x  \sqrt{\R^2}}{ \sqrt{\R^2}}\,,
\]
in analogy with the so-called \emph{quantum Lagrangian} \cite{Wyatt} for a free quantum particle, given by the last two terms above.  Then, upon  taking the total differential ${\rm d}$ (on $T^*Q\times M$) of \eqref{HybMad1a} and rewriting \eqref{HybHJR} in terms of the density $\R^2$, we may rewrite \eqref{HybHJ}-\eqref{HybHJR} as follows:
\begin{align}\label{HybMad1}
&\left(\frac{\partial }{\partial t}+\pounds_\mathsf{X}\right) {\rm d}{\cal S} =\de\mathscr{L}
\,,
\\
&\ \frac{\partial \R^2}{\partial t}+\operatorname{div}(\R^2 \mathsf{X}) =0\,.
\label{HybMad2}
\end{align}
In \eqref{HybMad2}, the operator $\operatorname{div}$ denotes the divergence operator induced on $T^*Q\times M$ by the Liouville form on $T^*Q$ and the Riemannian metric on $M$.
As we shall see, the above form of the hybrid Madelung equations will be crucially important.

Notice that, in the absence of coupling, we have $\partial_{q^j} \partial_{x^k}V=0$ in \eqref{QHam} and the mean-field factorization $\Upsilon(z,x)=\Psi_C(z)\Psi_Q(x)$ becomes an exact solution of \eqref{hybrid_KvH}. Indeed, in this case we have ${\cal S}(z,x)= S_C(z) +S_Q(x)$ and ${\cal R}(z,x)=R_C(z) R_Q(x)$, so that equation \eqref{HybMad1a} recovers the purely classical case \eqref{KvHMadelung1} for $S_C(z)$ and the purely quantum case \eqref{MadelungEqs} for $S_Q(x)$. Analogously, equation \eqref{HybMad2} leads to \eqref{KvHMadelung1} and \eqref{MadelungEqsbis} for $R_C(z)$ and $R_Q(x)$, respectively.

Before closing this section, we emphasize that the hybrid vector field  \eqref{HybridVecF} cannot be directly used to construct a probability current for the hybrid quantum--classical system. Indeed, while the vector field $\mathsf{X}\in\mathfrak{X}(\Gamma)$ transports the density $\mathcal{R}^2=|\Upsilon|^2$ appearing in \eqref{rhoq}, it does not transport the hybrid probability density, which instead must be constructed from the hybrid operator-valued density \eqref{hybridDenOp}. This topic is developed in the second part of the paper; see equation \eqref{Ddef} in Section \ref{sec:Dmomap}.

\subsection{Hybrid Bohmian trajectories \label{sec:Bohmian}}

As commented at the end of Section \ref{sec:equiv}, the absence of classical particle trajectories in hybrid dynamics raises the question whether a Hamiltonian flow can still be defined as incorporating the motion of the classical subsystem. In this section, a positive answer is provided in terms of Lagrangian paths extending quantum Bohmian trajectories to hybrid quantum--classical systems. For later use, it will be  convenient to introduce the shorthand notation $(z,x)\in\Gamma$, where
\[
\Gamma:=T^*Q\times M 
\] 
represents the hybrid quantum--classical coordinate manifold. Evidently, this is a volume manifold with volume form given by 
\begin{equation}\label{HvolForm}
\mu_\Gamma= \Lambda\wedge \mu\,.
\end{equation}

Although equation \eqref{HybMad1} is not in the typical hydrodynamic form, the hybrid Madelung equations \eqref{HybMad1}-\eqref{HybMad2} still lead to a similar continuum description to that obtained in the quantum case. For example, the density equation \eqref{HybMad2} still yields hybrid  trajectories, which may be defined by considering the evolution equation $\R^2(t)=(\R^2_0\circ \Phi(t)^{-1})/\sqrt{\operatorname{Jac}(\Phi(t))}$, where $\Phi(t)$ is the flow of the vector field $\mathsf{X}$ and $\operatorname{Jac}(\Phi)$ is the Jacobian relative to the volume form \eqref{HvolForm} on $\Gamma$. Then, this flow is regarded as a Lagrangian  trajectory obeying the equation
\begin{equation}\label{HybTraj}
\frac{\de}{\de t}\Phi(t,z,x)= \mathsf{X}(\Phi(t,z,x))
\,,
\end{equation}
which is the hybrid quantum--classical extension of the quantum Bohmian trajectories \cite{Bohm,Wyatt} in \eqref{QuantumPaths}. In turn, hybrid Bohmian trajectories \eqref{HybTraj} are also useful to express  \eqref{HybHJ} in the form
\[
\frac{\de}{\de t}{\cal S}(t,\Phi(t,z,x))= \mathscr{L}(t,\Phi(t,z,x))
\,,
\]
which is the hybrid analogue of the classical phase evolution \eqref{classicalphase_evolution}.  Additionally, in the absence of classical degrees of freedom, this picture recovers the quantum Bohmian trajectories since in that case the coordinate $z$ plays no role.

At this point, we address the question of how the symplectic property \eqref{simplecticity} of the flow is affected by quantum--classical coupling. In other words, here we shall unfold the geometric features underlying the dynamics of the classical canonical symplectic form  $\omega=\de q^i\wedge \de p_i$. 
By construction, we observe that the phase-space components $X_{H_I}$ of the hybrid vector field $\mathsf{X}$ in \eqref{HybridVecF} identify a Hamiltonian vector field parameterized by the coordinate $x\in M$, that is $X_{H_I}\in \mathcal{F}\big(M,\mathfrak{X}_\omega(T^*Q)\big)$. Indeed, upon denoting by $\de_z$ the differential on $T^*Q$, the relation $\mathbf{i}_{X_{H_I}}\omega={\rm d}_zH_I$ follows from a straightforward verification. Then, the flow $\tilde\eta$ of $X_{H_I}$ is symplectic so that
\beq\label{Amy}
\frac{\de}{\de t}\tilde\eta(t)_*\omega=0
\,.
\eeq
\noindent
Therefore, despite the absence of point particle trajectories in quantum--classical coupling, a Hamiltonian flow preserving the classical symplectic form can still be defined as the flow associated to the Hamiltonian $H_I$. Notice that ${\tilde\eta(t)\in \mathcal{F}\big(M,  \operatorname{Diff}_\omega(T^*Q)}\big)$ differs from the Lagrangian trajectory $\Phi (t) \in \operatorname{Diff}(T^*Q\times M)$ of the hybrid system \eqref{HybTraj}, unless the quantum kinetic energy operator $-({\hbar^2}/{2m})\Delta_x$ is dropped from the hybrid Hamiltonian \eqref{QHam}. Indeed, in the latter case the hybrid vector field \eqref{HybridVecF} drops to $\mathsf{X}=(X_{H_I},0)$ and one is left with
$\Phi(t)(z,x)=( \tilde\eta(t)(z,x), x)$,  so that the hybrid Lagrangian trajectory $\Phi(t)$ is essentially equivalent to the path $\tilde\eta(t)$.

\subsection{Hybrid dynamics and symplectic form\label{sec:sympform}}
When the quantum kinetic energy is retained, the classical symplectic form is not preserved in the whole hybrid space $T^*Q\times M$ and here we shall present its corresponding dynamics. In particular, we have the following result:
\begin{theorem}[Hybrid trajectories and symplectic form] Let $\pi_{T^*Q}:\Gamma \rightarrow T^*Q$ be the standard projection map and let $\mathsf{X}\in\mathfrak{X}(\Gamma)$ be the hybrid vector field  \eqref{HybridVecF}, with $H_I$ given by \eqref{intHam}. Also, denote by $\Phi\in \operatorname{Diff}(\Gamma)$  the flow of $\mathsf{X}$ as in \eqref{HybTraj} and define the two-form $\Omega:=\pi^*_{T^*Q\,}\omega\in \Omega^2(\Gamma)$  naturally induced on the hybrid coordinate space $\Gamma$. Then, we have the statements below.
\begin{enumerate}

\item The mapping $X_{H_I}:\Gamma\to TT^*Q$ given by $X_{H_I}=T\pi_{T^*Q}\circ{\sf X}$ defines a parameterized Hamiltonian vector field, i.e. $X_{H_I}\in \mathcal{F}\big(M,\mathfrak{X}_\omega(T^*Q)\big)$;

\item the map $\tilde\eta:\Gamma \to T^*Q$ given  by $\tilde\eta:=\Phi^*\pi_{T^*Q}$ defines a parameterized symplectic diffeomorphism, i.e. $\tilde\eta\in\mathcal{F}\big(M,\operatorname{Diff}_\omega(T^*Q)\big)$, which identifies the flow of $X_{H_I}$;

\item the following relation holds:
\beq\label{sympformdyn}
\frac{\de }{\de t}\Phi(t)^*\Omega= -{\rm d}\big(\Phi(t)^* \pi_M^*{\rm d}_x V\big)\,,\quad\ \text{with}\quad\  {\rm d} (\pi_M^*{\rm d}_x V)= \frac{\partial^2 V}{\partial q^{j} \partial x^k}\,{\rm d} q^j\wedge {\rm d}x^k\,.
\eeq
\end{enumerate}
\end{theorem}
\paragraph{Proof.} The first two statement follow from the discussion at the end of the previous section. For the third point, if $\mathsf{X}\in\mathfrak{X}(\Gamma)$ is the hybrid vector field \eqref{HybridVecF} and $\pi_{T^*Q}:\Gamma \rightarrow T^*Q$, $\pi_{M}:\Gamma \rightarrow M$ are the projection maps, then we may consider the one-form $\mathsf{A}:=\pi^*_{T^*Q}{\mathcal{A}}\in \Omega^1(\Gamma)$ naturally induced by the canonical one-form $\mathcal{A}$ on $T^*Q$ and we have
\[
\pounds_\mathsf{X}\mathsf{A} = {\rm d} L_I + \pi^*_{M}{\rm d}_x V,
\]
where  $\de_x$ denotes the differential on $M$. This relation is obtained by a direct verification as follows:
\begin{align*}
\pounds_\mathsf{X} \mathsf{A} =&\ 
 \mathbf{i}_\mathsf{X}{\rm d} \mathsf{A} +{\rm d}(\mathbf{i}_\mathsf{X} \mathsf{A})
 \\
=  &\ \mathbf{i}_\mathsf{X}{\rm d}\pi^*_{T^*Q}{\mathcal{A}} +{\rm d}(\mathbf{i}_\mathsf{X}\pi^*_{T^*Q}{\mathcal{A}}) 
 \\
= &   -\mathbf{i}_\mathsf{X}\pi^*_{T^*Q}\omega +{\rm d}(\mathbf{i}_{X_{H_I}}{\mathcal{A}})\\
=&- \pi^*_{T^*Q}{\rm d}_z H_I +{\rm d}(\mathbf{i}_{X_{H_I}}{\mathcal{A}}) 
\\
=& - {\rm d}H_I +\pi_M^* {\rm d}_x H_I + {\rm d}(\mathbf{i}_{X_{H_I}}{\mathcal{A}})
\\
=
&\ 
{\rm d} L_I + \pi_M^*{\rm d}_x V.
\end{align*}
Inserting this relation in \eqref{HybMad1} yields
\begin{equation}\label{Important}
\left(\partial_t+\pounds_\mathsf{X}\right) (\de {\cal S} - \mathsf{A} ) =\frac{1}{2m} \de\bigg(|\nabla_x {\cal S}|^2+ \hbar^2\frac{\Delta_x \sqrt{D}}{ \sqrt{D}}\bigg) - \pi_M^* {\rm d}_xV
\,,
\end{equation}
which is equivalently written as
\begin{equation}\label{A_dynamics}
\frac{\de }{\de t}\Phi(t)^*\big(\de {\cal S} - \mathsf{A} \big)=\frac{1}{2m} \de \big( \Phi(t)^* (\mathscr{L}-L_I)\big) - \Phi(t)^* \pi_M ^* {\rm d}_xV\,.
\end{equation}
Then, taking the exterior differential $\de$ on $\Gamma$ of the relation \eqref{A_dynamics} yields \eqref{sympformdyn}. \qquad\qquad$\blacksquare$

\medskip

\noindent
The relation \eqref{sympformdyn} describes the evolution of the classical canonical form under the whole hybrid flow $\Phi$. As expected, the canonical symplectic form is not preserved by this flow unless the quantum--classical coupling vanishes, that is $\partial^2_{q^{j\scriptscriptstyle\!} x^k} V=0$. Notice that, in local coordinates, $\mathsf{A}=p_i\de q^i$ and $\pi_M ^* {\rm d}_xV=\partial_{x^i} V\de x^i$, so that \eqref{A_dynamics} produces the following equation for the Poincar\'e integral in the hybrid coordinate space $\Gamma$:
\[
\frac{\de }{\de t}\oint_{\gamma(t)}p_i\,\de q^i = -\oint_{\gamma(t)} \frac{\partial V}{\partial x^i} \, \de x^i
\,
\]
where  $\gamma(t)=\Phi(t)\circ\gamma_0$ and $\gamma_0$ is a time-independent loop in $\Gamma$. 

In summary, we have found that, although the classical canonical form is not preserved by the overall hybrid flow $\Phi(t)$ of $\mathsf{X}\in\mathfrak{X}(\Gamma)$, it is preserved by the Hamiltonian flow $\tilde\eta(t)\in \mathcal{F}\big(M,\operatorname{Diff}_\omega(T^*Q)\big)$ of $X_{H_I}\in \mathcal{F}\big(M, \mathfrak{X}(T^*Q)\big)$, given by the first component of the hybrid flow $\Phi(t)(z,x)=(\tilde\eta(t)(z,x), \zeta(t)(z,x))$. In turn, the flow $\tilde\eta(t)$ determines an important subgroup of  the group $\operatorname{Diff}(\Gamma)\,\circledS \,\mathcal{F}(\Gamma, S^1)$ advancing the full hybrid dynamics.
\begin{lemma}[A subgroup of the hybrid flow]\label{subgrouplemma} We have the following group inclusion
\begin{equation}\label{subgroup_inclusion} 
\mathcal{F}\big(M,\widehat{\operatorname{Diff}}_\omega(T^*Q)\big) \subset\operatorname{Diff}(\Gamma)\,\circledS \,\mathcal{F}(\Gamma, S^1), \qquad \qquad(\eta, e^{{\rm  i} \theta})
\mapsto 
\left(\tilde\eta,e^{{\rm i}\theta+{\rm i}\int_{z_0}^{z}\left({\mathcal{A}} -\eta^*{\mathcal{A}}\right)}\right)
\end{equation} 
where $ \eta \in \mathcal{F} (M, \operatorname{Diff}_\omega(T^*Q))$, $ \theta \in \mathcal{F} (M, S^1)$, and with $\tilde \eta \in \operatorname{Diff}( \Gamma )$ defined by  $\tilde{ \eta }(z,x)=(\eta (x)(z), x)$.
The associated Lie algebra inclusion reads
\begin{eqnarray}\nonumber
\iota: \mathcal{F} ( \Gamma )\,  \simeq&\!\!\! \mathcal{F} \big(M, \widehat{ \mathfrak{X} }_ \omega  (T^*Q)\big) &\!\!\!\hookrightarrow \mathfrak{X}(\Gamma)\,\circledS\, \mathcal{F}(\Gamma) 
\\
 \xi  \, \simeq&\!\!\!  (X_\xi , - \xi _ \mathcal{A}(z_0))  
 &\!\!\!\mapsto \big((X_\xi,0), -\xi_{\mathcal{A}}\big).
 \label{sub_Lie_algebra_inclusion} 
\end{eqnarray} 
\end{lemma}
\paragraph{Proof.} The group inclusion is more naturally written by inserting an intermediate subgroup as follows
\[
\mathcal{F}\big(M,\widehat{\operatorname{Diff}}_\omega(T^*Q)\big) \subset \mathcal{F} \big( M,\operatorname{Diff}(T^*Q)\,\circledS \,\mathcal{F}(T^*Q, S^1)\big) \subset \operatorname{Diff}(\Gamma)\,\circledS \,\mathcal{F}(\Gamma, S^1).
\]
The first inclusion is associated to the group inclusion $\widehat{\operatorname{Diff}}_\omega(T^*Q) \subset \operatorname{Diff}(T^*Q)\,\circledS \,\mathcal{F}(T^*Q, S^1)$, which follows from \eqref{stricts} and \eqref{groupisomorphism}. The second inclusion is $( \eta , f) \mapsto (\tilde{ \eta }, \tilde{ f})$ with $\tilde{ \eta }(x,z)=(x, \eta (x)(z))$ and $\tilde f(x,z)= f(x)(z)$, for $\eta(x) \in \operatorname{Diff}(T^*Q)$ and $f(x) \in \mathcal{F} (T^*Q, S^1)$, which can be checked to be a group homomorphism. The composition of these group inclusions yields \eqref{subgroup_inclusion}. 

The Lie algebra isomorphism $\mathcal{F} ( \Gamma ) \simeq \mathcal{F} \big(M, \widehat{ \mathfrak{X} }_ \omega  (T^*Q)\big)$ given by $\xi  \simeq  (X_\xi , - \xi _ \mathcal{A}(z_0)) $ follows from the Lie algebra isomorphism \eqref{LAisomorphism}, where we identify $ \mathcal{F} ( \Gamma )$ with  $\mathcal{F} (M, \mathcal{F} (T^*Q))$. Taking the derivative of \eqref{subgroup_inclusion} at the identity yields the Lie algebra inclusion $(X_ \xi , \kappa ) \in   \mathcal{F} \big(M, \widehat{ \mathfrak{X} }_ \omega  (T^*Q)\big)\mapsto ((X_ \xi , 0), \kappa + \xi _\mathcal{A}(z_0) - \xi _ \mathcal{A}) \in \mathfrak{X} ( \Gamma ) \,\circledS\, \mathcal{F} ( \Gamma )$ similarly as computed in 
\eqref{Lie_algebra_isom_aut}. By composing with the Lie algebra isomorphism $\mathcal{F} ( \Gamma ) \simeq \mathcal{F} \big(M, \widehat{ \mathfrak{X} }_ \omega  (T^*Q)\big)$ we get the desired formula.
$\qquad\qquad\blacksquare$

\medskip \noindent
Section \S\ref{Sec_probability} will show that the group $\mathcal{F}\big(M,\widehat{\operatorname{Diff}}_\omega(T^*Q)\big)$ is of fundamental importance in the probabilistic interpretation of the hybrid quantum--classical wave equation \eqref{CQwaveq}. 

In the next section, we shall study the hybrid Madelung equations \eqref{HybMad1}-\eqref{HybMad2} in terms of its Hamiltonian  structure, which arises naturally from the momentum map property of the Madelung transform \cite{FoHoTr}.

\subsection{Hamiltonian structure and equations of motion}\label{Ham_var}

The equations \eqref{HybMad1}-\eqref{HybMad2} possess a Hamiltonian structure whose Lie-Poisson bracket is identical to that governing compressible barotropic fluids \cite{MaRaWe}. This is due to the fact that the mapping
\begin{eqnarray}\nonumber
J:\; L^2(\Gamma)&\!\!\to&\!\!\mathfrak{X}(\Gamma)^*\times \operatorname{Den}(\Gamma)
\\
\Upsilon&\!\!\mapsto&\!\!J(\Upsilon)=\big(\hbar\operatorname{Im}(\bar \Upsilon{\rm d}\Upsilon),|\Upsilon|^2\big)=
\big(\R^2{\rm d}{\cal S},\,\R^2\big)
\label{Madelungmomap}
\end{eqnarray}
comprise a momentum map structure that is the hybrid analogue of its quantum and classical variants given by \eqref{QuantumMadelungMomap} and \eqref{Joli}, respectively. Here, $\operatorname{Den}(\Gamma)$ and $\mathfrak{X}(\Gamma)^*$ denote respectively the space of densities and one-form densities on $\Gamma=T^*Q\times M$. Upon identifying  $L^2(\Gamma,\mathbb{C})$  with the space of half-densities, the momentum map \eqref{Madelungmomap} is produced by the (left) representation
\begin{equation}\label{Hybrid_Madelung_action}
\Upsilon\mapsto \frac1{\sqrt{ \operatorname{Jac}(\Phi) }}\ \Phi_*\big(e^{-i{\varphi}/\hbar}\Upsilon\big)
\end{equation}
of the semidirect-product group $\operatorname{Diff}(\Gamma)\,\circledS \,\mathcal{F}(\Gamma)\ni(\Phi,\varphi)$.

\begin{figure}[h]
\footnotesize\center
\begin{framed}
\begin{xy}
\hspace{.4cm}
\xymatrix{
&
*+[F-:<3pt>]{
\begin{array}{l}
\vspace{0.1cm}\text{Hybrid wave equation \eqref{hybrid_KvH}}\\
\vspace{0.1cm}\Upsilon \in \mathscr{H}_{\scriptscriptstyle C Q}\\
\vspace{0.1cm}\displaystyle h(\Upsilon)= \int_{T^*Q}\!\big\langle\Upsilon\big|\widehat{\cal L}_{\widehat{H}}\Upsilon\big\rangle\,\Lambda\\
{\rm i}\hbar \partial _t \Upsilon = \widehat{\cal L}_{\widehat{H}}\Upsilon
\end{array}} \ar[ddrrr]|{\begin{array}{c}\text{Momentum map \eqref{Madelungmomap}}\\
\text{ for the group}\\
\operatorname{Diff}(\Gamma) \,\circledS\, \mathcal{F}(\Gamma, S^1)\\
\end{array}}
\ar[rrr]|{}
\ar[dd]|{\begin{array}{l}\Upsilon= \mathcal{R} e^{{\rm i}  \mathcal{S} /\hbar}\end{array}} & & &*+[F-:<3pt>]{
\begin{array}{l}
\vspace{0.1cm}\text{Hybrid density operator \eqref{hybridDenOp}}\\
\vspace{0.1cm} \widehat{ \mathcal{D}}  \in \operatorname{Den}(T^*Q, \operatorname{Her}(\mathscr{H}_{\scriptscriptstyle Q}))\\
\vspace{0.1cm} \text{Hamiltonian $h( \widehat{ \mathcal{D} })$}
\end{array}
}\ar@{--}[dd]|{\begin{array}{l}\text{See Section \ref{Sec_probability}}\end{array}} \\
& & & & &\\
&*+[F-:<3pt>]{\begin{array}{l}\text{Hybrid Madelung}\\
\text{equations \eqref{HybHJ}-\eqref{HybHJR} }\\
\text{for $( \mathcal{S} , \mathcal{R} )$}
\end{array}}\ar[rrr] & & &
*+[F-:<3pt>]{
\begin{array}{l}
\vspace{0.1cm}\text{Hybrid Madelung variables}\\
\vspace{0.1cm} ( \sigma , D)   \in \big( \mathfrak{X}(\Gamma) \,\circledS\, \mathcal{F}(\Gamma) \big)^*\\ 
\vspace{0.1cm} \text{Hamiltonian $h( \sigma , D)$ in \eqref{collective2}}\\ 
\vspace{0.1cm} \text{Hamiltonian system \eqref{LPEqns1}}  
\end{array}
} 
&
}
\end{xy}
\end{framed}
\vspace{-.5cm}\it
\caption{Schematic description of the various quantities appearing in quantum--classical mechanics and some of the mappings between them. The quantities in the boxes on the left will be related in Section \ref{Sec_probability}.}
\label{figure2}
\end{figure}
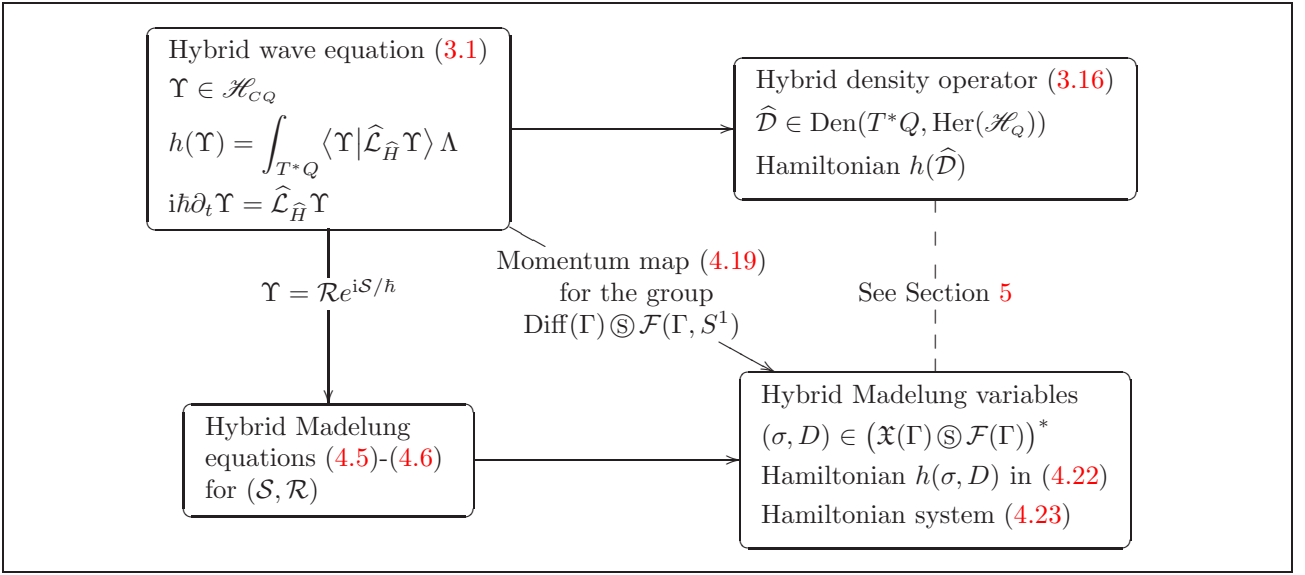

Since the momentum map \eqref{Madelungmomap} is equivariant, it is also a Poisson map, thereby producing the  Lie-Poisson structure on the dual of the semidirect-product Lie algebra $\mathfrak{X}(\Gamma)\,\circledS\, \mathcal{F}(\Gamma)$. More explicitly, upon defining the one form $\sigma=\R^2{\rm d}{\cal S}$ and the density $D=\R^2$ on $\Gamma$, the Lie-Poisson bracket reads
 \begin{align}
\{f\,,\,h\}_{\scriptscriptstyle \rm LP}(\sigma,D) \nonumber
= - \int _\Gamma\!\bigg[&\, \sigma_j \left( \frac{\delta f}{\delta {\sigma}} \cdot \nabla  \frac{\delta h}{\delta \sigma_j}
- \frac{\delta h}{\delta {\sigma}} \cdot \nabla  \frac{\delta f}{\delta \sigma_j} \right)
\\
&\qquad\qquad
+ D \left( \frac{\delta f}{\delta {\sigma}} \cdot \nabla  \frac{\delta h}{\delta D}
- \frac{\delta h}{\delta {\sigma}} \cdot \nabla  \frac{\delta f}{\delta D} \right)\!\bigg] \mu_\Gamma
\,.
\label{LPB}
\end{align}
With the above bracket, the Hamiltonian functional producing the Madelung equations \eqref{HybMad1}-\eqref{HybMad2} is
\begin{align}
h(\sigma,D)=
\int_\Gamma\left(\frac{1}{2m}\frac{|\sigma_x|^2}{D} + \frac{\hbar^2}{8m}\frac{|\nabla_x D|^2}{D}
-D L_I+ \sigma_z\cdot X_{H_I}\right) \mu_\Gamma
\,, \label{collective2}
  \end{align}
where we have used the notation
\[
\sigma=\R^2{\rm d}{\cal S}=\left(\R^2{\rm d}_z{\cal S}\,,\R^2{\rm d}_x{\cal S}\right)=:(\sigma_z,\sigma_x)
\]
to split classical and quantum  components. Here, the symbols $\de_z$ and $\de_x$ denote the exterior differentials on $T^*Q$ and $M$, respectively. The Hamiltonian \eqref{collective2} is obtained by rewriting the Hamiltonian \eqref{hybHam}  in  terms of $\sigma$ and $D$. Then, equations \eqref{HybMad1}-\eqref{HybMad2} are obtained from a direct verification via the Poisson bracket formulation $\de{f}/\de t=\{f,h\}_{\scriptscriptstyle \rm LP}$ and by writing $D={\cal R}^2$ and $\de{\cal S}=\sigma/D$.
The Hamiltonian structure   \eqref{LPB}-\eqref{collective2} yields the equations \eqref{HybMad1}-\eqref{HybMad2} in the following Lie-Poisson form:
\beq\label{LPEqns1}
\left(\frac{\partial }{\partial t}+\pounds_\mathsf{X}\right) \sigma =D\de\mathscr{L}
\,,\qquad\qquad
\frac{\partial D}{\partial t}+\operatorname{div}(D \mathsf{X}) =0
\,.
\eeq
The general picture summarizing the various steps in hybrid quantum--classical mechanics is shown in Fig. \ref{figure2}.

\smallskip

At this point, it may be interesting to project equation \eqref{HybMad1} for  $\de{\cal S}=\sigma/D$ onto both its phase-space and quantum components. This task can be quickly achieved by taking differentials of equation \eqref{HybMad1a}. First, we consider the phase function ${\cal S}$ in \eqref{polarform} as a function on $T^*Q$ parameterized by the quantum coordinate, that is, instead of $\mathcal{S}\in\mathcal{F}(T^*Q\times M)$, we interpret the phase as a function 
${\cal S}\in \mathcal{F}\big(T^*Q,\mathcal{F}(M)\big)$.
Then, defining ${\cal V}_Q:=\nabla_x S/m\in \mathcal{F}\big(T^*Q,\mathfrak{X}(M)\big)$ and using the Lie derivative $\pounds_{{\cal V}_Q}$  on $M$ leads to rewriting \eqref{HybMad1a} in the equivalent form
\[
\left(\frac{\partial }{\partial t}+\pounds_{{\cal V}_Q\!}\right) {\cal S} =\mathscr{L}+\{H_I,{\cal S}\}
\,,
\]
so that taking the differential $\de_x$ yields
$
\left(\partial_t+\pounds_{{\cal V}_Q}\right) \de_x{\cal S} =\de_x(\mathscr{L}-{X_{H_I}}\cdot\nabla_z{\cal S})
$.
At a second stage, we may proceed alternatively by setting
$
{\cal S}\in \mathcal{F}\big(M,\mathcal{F}(T^*Q)\big)
$.
Then,   defining 
\[
\mathcal{H}_Q:=\frac{|\nabla_x  {\cal S}|^2}{2m} - \frac{\hbar^2}{2m}\frac{\Delta_x  \sqrt{D}}{ \sqrt{D}}\, \in \mathcal{F}\big(M,\mathcal{F}(T^*Q)\big)
\]
takes equation  \eqref{HybMad1a} into the equivalent form
\[
\left(\frac{\partial }{\partial t}+\pounds_{X_{H_I}\!}\right)\mathcal{S}+\mathcal{H}_Q= L_I
\,,
\]
where we recall that  $X_{H_I}\in \mathcal{F}\big(M,\mathfrak{X}(T^*Q)\big)$ and $\pounds_{X_{H_I}}$ is the Lie derivative on $T^*Q$.
Then, taking the differential $\de_z$ of the above yields
$
(\partial_t+ \pounds_{X_{H_I}})\de_z\mathcal{S}=\de_z (L_I-\mathcal{H}_Q)
$. 
Then, the Lie-Poisson equations \eqref{LPEqns1} are equivalently written as
\begin{align}\label{Valkiria1}
&\,\left(\frac{\partial }{\partial t}+\pounds_{{\cal V}_Q\!}\right) \de_x{\cal S} =\de_x(\mathscr{L}-{X_{H_I}}\cdot\nabla_z{\cal S})
\,,\\\label{mario}
&\,\left(\frac{\partial }{\partial t}+ \pounds_{X_{H_I}\!}\right)\de_z\mathcal{S}= \de_z (L_I-\mathcal{H}_Q)
\,,\\
&\, \left(\frac{\partial }{\partial t}+\pounds_{\sf X}\right)D =0
\,.
\end{align}
Despite its similarities with standard quantum hydrodynamics, equation \eqref{Valkiria1}  differs from the first in \eqref{MadelungEqs2} in that it retains the important term ${X_{H_I}}\cdot\nabla_z{\cal S}$, which persists in the absence of quantum--classical coupling. This is due to the fact that the function ${\cal S}(t,z,x)$ is not the quantum phase, but rather it is a phase-like quantity associated to the compound quantum--classical system evolving along the hybrid Lagrangian trajectory $\Phi$  with Eulerian velocity $\mathsf{X}= (X_{H_I},\nabla_x{\cal S}/m)$.  
In order to understand the information carried by equation \eqref{mario}, we are going to rewrite it in a more convenient form.
Similarly to above, we may redefine the canonical one-form ${\mathcal{A}}=p_i{\rm d} q^i$ as a one-form on $T^*Q$ parameterized by the quantum coordinate space, that is ${\mathcal{A}} \in\mathcal{F}\big(M,\Omega^1(T^*Q)\big)$, and we observe that
$\pounds_{X_{H_I}}{\mathcal{A}} = {\rm d}_z L_I$ so that $X_{H_I}$ is Hamiltonian, i.e. $X_{H_I}\in\mathcal{F}\big(M,\mathfrak{X}_\omega(T^*Q)\big)$. Consequently,  equation \eqref{mario} may also be rewritten as
\begin{equation}
\left(\frac{\partial }{\partial t}+ \pounds_{X_{H_I}\!}\right) ({\rm d}_z {\cal S}-{\mathcal{A}}) =-{\rm d}_z\mathcal{H}_Q
\end{equation}
which implies
$\big({\partial_t}+\pounds_{X_{H_I}}\big) {\rm d}_z{\mathcal{A}} =0$, thereby recovering \eqref{Amy} for the flow $\tilde\eta(t)$ of $X_{H_I}$.

While the quantum--classical Madelung transform has been characterized in its geometric content and related to hybrid Bohmian trajectories, the second part of this paper considers the role of the Madelung transform in the context of probability densities. The latter will be studied in terms of their continuity equation and the associated quantum--classical currents. As we shall see, this study will establish a relation between the two boxes on the \makebox{right in Fig.~\ref{figure2}.}


\section{Hybrid quantum--classical densities and currents}\label{Sec_probability}

While the previous sections presented the main geometric properties of the hybrid Madelung equations \eqref{HybMad1}-\eqref{HybMad2}, here we want to focus on their physical interpretation in terms of probability densities and  currents. 

\subsection{General comments}
As presented in Section \ref{sec:CQeq}, the general hybrid density associated to the quantum--classical wavefunction in \eqref{hybrid_KvH} is given by the operator-valued distribution $\widehat{\cal D}(z)$ in \eqref{hybridDenOp}. At present, there is no criterion available to establish whether the dynamics of $\widehat{\cal D}(z)$ preserves its sign  \cite{BoGBTr}, unless one considers the trivial case in which the quantum--classical interaction is absent. Indeed, in the latter case, the hybrid wave equation \eqref{hybrid_KvH} with $\widehat{H}=\widehat{H}_Q+H_C$ produces the following evolution equation for $\widehat{\cal D}$:
\begin{equation}\label{DeqNocoup}
\partial_t\widehat{\cal D}=-{\rm i}\hbar^{-1}[\widehat{H}_Q,\widehat{\cal D}]+\{H_C,\widehat{\cal D}\}\,.
\end{equation}
Here, we have assumed  a  potential $V(q,x)=V_Q(x)+V_C(q)$ in  \eqref{QHam}  so that the hybrid Hamiltonian is written as $\widehat{H}(q,p)=\widehat{H}_Q+H_C(q,p)$, where the subscripts $Q$ and $C$ refer respectively to quantum and classical. It is obvious that  the evolution \eqref{DeqNocoup} preserves the sign of $\widehat{\cal D}$, which then remains positive-definite in time. Indeed, upon using the notation in \eqref{cov2}-\eqref{cov1}, equation \eqref{DeqNocoup} leads to the sign-preserving evolution $\widehat{\cal D}(t)=\widehat{U}(t)(\eta(t)_*\widehat{\cal D}_0)\widehat{U}(t)^\dagger$, with ${\rm i}\hbar\,\de{\widehat{U}}/\de t=\widehat{H}_Q\widehat{U}$ and $\de{\eta}/\de t=X_{\scriptscriptstyle H_C}\circ\eta(t)^{-1}$.

 However, in the general case the equation of motion for $\widehat{\cal D}$  is sensibly more complicated  as it involves the hybrid wavefunction $\Upsilon$ as well as its gradients \cite{BoGBTr}. Then, the study of the evolution of the sign of $\widehat{\cal D}$ becomes very challenging. So far, all we know is that currently the hybrid  theory in Section \ref{sec:CQeq} is  the only Hamiltonian quantum--classical correlation theory  that 1) retains the quantum uncertainty principle (since \eqref{QuantDensMat} is positive definite) and 2) allows the mean-field factorization as an exact solution in the absence the quantum--classical coupling.
On the other hand, no  statement is yet available on the sign of the classical density $\rho_c(z)=\operatorname{Tr}(\widehat{\cal D}(z))$ and one is led to consider the possibility that  $\rho_c$ may assume negative values. Following Feynman's work  \cite{Feynman}, this point was justified in \cite{BoGBTr} by using arguments involving the Wigner function for a harmonic oscillator coupled to a nonlinear quantum system: even in that simple case, the oscillator distribution must be allowed to acquire negative values. Still, in \cite{BoGBTr} an example of hybrid dynamics was provided in which the classical density remains positive. Then, the question arises of characterizing possible cases in which the classical positivity is preserved in time.

\subsection{Hybrid density function as a momentum map}\label{sec:Dmomap}

Instead of considering the evolution of the classical density $\rho_c$ as it arises from the $\widehat{\cal D}-$equation, the remainder of this paper focuses on the dynamics of the diagonal elements ${\cal K}_{\widehat{\cal D}}(z;x,x)$ of the kernel \eqref{kappa_D} of $\widehat{{\cal D}}$, which we  denote as 
\begin{equation}\label{Ddef}
{\cal D}(z,x):={\cal K}_{\widehat{\cal D}}(z;x,x)= |\Upsilon(z,x)|^2-  \partial_{p_i} \big(p_i|\Upsilon(z,x)|^2\big) + {\rm i}\hbar \{\Upsilon, \bar\Upsilon\}(z,x)
\,.
\end{equation}
Notice that the mapping  $\widehat{\cal D}\mapsto{\cal D}$ is given by the dual of the vector space inclusion $\mathcal{F} ( \Gamma ) \hookrightarrow \mathcal{F} (T^*Q, \operatorname{Her}(\mathscr{H}_{\scriptscriptstyle Q}))$ given by $\xi(z,x)\mapsto\,\widehat{\!\xi}(z)$ and underlying the hybrid Hamiltonian $H_I(z,x)$.
In terms of the polar form of the hybrid wavefunction \eqref{polarform}, one has
\begin{equation}\label{polarD}
{\cal D}= \R^2+\partial_{p_i}(p_i\R^2)+\{\R^2,{\cal S}\}
\,.
\end{equation}
This quantity represents a joint density function for the position of the system in the hybrid coordinate space $T^*Q\times M$, in such a way that the quantum and the classical probabilities defined in \eqref{rhoq} and \eqref{rhoc} can be computed from ${\cal D}$ as 
\beq\rho_q(x)=\int_{T^*Q}{\cal D}(z,x)
\,\Lambda
\qquad\quad\text{and}\qquad\quad  \rho_c(z)=\int_M{\cal D}(z,x)\,\mu
\,.\label{luigi}
\eeq
Thus, finding the evolution equation of $\cal D$ allows characterizing a hybrid current ${\bf J}$ such that $\partial_t{\cal D}=-\operatorname{div}{\bf J}$. 
While this will be the subject of the next section, here we show how the quantity \eqref{Ddef} is actually a momentum map for the action of the group $\mathcal{F}\big(M,\widehat{\operatorname{Diff}}_\omega(T^*Q)\big)$ on the space $\mathscr{H}_{\scriptscriptstyle QC}=L^2(T^*Q\times M)$ of hybrid wavefunctions. 

Given an element $(\tilde\eta(x),e^{{\rm i}\kappa(x)})\in \mathcal{F}\big(M,\widehat{\operatorname{Diff}}_\omega(T^*Q)\big)$, its (left) action on $\Upsilon\in \mathscr{H}_{\scriptscriptstyle QC}$ can be constructed by suitably adapting the the van Hove representation \eqref{cl_prop} as follows:
\begin{equation}\label{action_M_SC}
\Upsilon(z,x)\mapsto \Upsilon\big( \tilde\eta^{-1}(z;x),x\big)\exp\left[-\frac{\rm i}{\hbar}\left(\kappa(x)+\int_{\tilde\eta(z_0;x)}^{z}(\tilde\eta(x)_*{\mathcal{A}} - {\mathcal{A}})\right)\right]\,.
\end{equation}
Here, the notation is such that $\tilde\eta(x)$ identifies a symplectic diffeomorphism $z\mapsto\tilde\eta(x)(z)=\tilde\eta(z;x)$. We shall drop the explicit dependence on the phase-space coordinates where convenient. Then, the KvH construction summarized in Section \ref{sec:KvH} is naturally transferred to the case of parameterized transformations: the Lie algebra of $\mathcal{F}\big(M,\widehat{\operatorname{Diff}}_\omega(T^*Q)\big)$ is the space $\mathcal{F}\big(M,C^\infty(T^*Q))\simeq \mathcal{F}(T^*Q\times M)$ of $x-$dependent phase-space Hamiltonians endowed with the canonical Poisson bracket on $T^*Q$, the dual space $\operatorname{Den}(T^*Q\times M)$ is the space of joint distributions on the hybrid space $T^*Q\times M$, and the infinitesimal generator of a parameterized Hamiltonian function $\xi(z,x)\in \mathcal{F}\big(M,C^\infty(T^*Q)\big)$ is $\Upsilon\mapsto -{{\rm i}}\hbar ^{-1}\widehat{\cal L}_{\xi(x)}\Upsilon$, see  Section \ref{central_ext_sym}. The momentum map for the action \eqref{action_M_SC} is found from the relation
\begin{equation}\label{MomapFormula_hybrid}
\langle {\cal D}(\Upsilon),\xi\rangle=\frac{1}{2}\Omega\big(-\rm i\hbar ^{-1}\widehat{\cal L}_{\xi}\Upsilon,\Upsilon\big)
\,,
\end{equation}
where $\langle\ ,\,\rangle$ denotes the duality pairing between $\mathcal{F}(T^*Q\times M)$ and its dual $\operatorname{Den}(T^*Q\times M)$ and the symplectic form is given by $\Omega(\Upsilon_1,\Upsilon_2)=2\hbar\operatorname{Im}\int_\Gamma\bar \Upsilon_1\Upsilon_2 \mu_\Gamma$. Then, \eqref{MomapFormula_hybrid} leads to the  momentum map 
\begin{equation}\label{Dmomap}
{\cal D}(\Upsilon)= |\Upsilon|^2+\partial_{p_i}(p_i|\Upsilon|^2)+{\rm i}\hbar\{\Upsilon,\bar\Upsilon\}
\,,
\end{equation}
which is the natural extension of expression \eqref{KvHmomap} and recovers precisely \eqref{Ddef}.

The momentum map structure of the hybrid  density $\cal D$ provides much insight into the geometry of its evolution. For example, dropping the quantum kinetic energy operator $-({\hbar^2}/{2m})\Delta_x$ from the hybrid Hamiltonian \eqref{QHam} produces a classical Liouville equation parameterized by $x$, that is
\begin{equation}
\partial_t{\cal D}=\{H_I,{\cal D}\}
\,,
\label{Deq1}
\end{equation}
which can be deduced from \eqref{polarD} by using that \eqref{HybHJ} and \eqref{HybMad2} reduce to $\partial_t  {\cal S}= L_I+\left\{H_I, {\cal S}\right\}$ and $\partial_t \R ^2= \left\{H_I,  \R^2\right\}$, respectively. Equation \eqref{Deq1} does not come as a surprise, since we already observed in Section \ref{sec:Bohmian} that dropping the quantum kinetic energy makes the hybrid Lagrangian path $\Phi$ coincide with the flow of the $x-$dependent Hamiltonian vector field $X_{H_I}$.  
In turn,  given the characteristic nature of equation \eqref{Deq1}, the latter possesses Klimontovich-like solutions of the form
\beq
{\cal D}(z,x,t)=w(x)\delta(z-\zeta(x,t))
\,,
\label{michele}
\eeq
where $w(x)\in \operatorname{Den}(M)$ and $\de\zeta/\de t=X_{H_I}(\zeta)$. Then, the classical Liouville density \eqref{luigi} reads
\[
\rho_c(z,t)=\int_M\! w(x)\delta(z-\zeta(x,t))\,\mu
\,.
\]
As shown in \cite{HoTr}, this  expression of the classical Liouville density identifies the left leg of a dual pair of momentum maps \cite{Weinstein}.  

Furthermore, when the quantum kinetic energy is drpped in \eqref{QHam},  a possibly relevant consequence of equation \eqref{Deq1} is that the sign of the joint probability density $\cal D$ is preserved in time even if the same conclusion cannot be generally reached about the operator-valued density $\widehat{\cal D}$. Evidently, the sign of $\cal D$ is also preserved in the absence of quantum--classical coupling, that is when $\partial^2_{q^{j\scriptscriptstyle\!} x^k} V_I=0$. In this trivial case,  equation \eqref{DeqNocoup} preserves the sign of $\widehat{\cal D}$ and therefore also the sign of its diagonal elements. At present, similar statements about sign conservation
are unavailable in the more general case of the hybrid Hamiltonian \eqref{QHam}. However, it may still be interesting to write down the continuity equation for $\cal D$ in order to characterize the corresponding quantum--classical current. This is the focus of the next section.

We conclude this section by extending the discussion at the end of Section \ref{sec:KvHMadelung} to the case of hybrid quantum--classical dynamics. In analogy to the representations \eqref{sdp-action} and  \eqref{cl_prop} and in agreement with Lemma \ref{subgrouplemma}, the representation \eqref{Hybrid_Madelung_action} of $\operatorname{Diff}(\Gamma )\,\circledS\,\mathcal{F}(\Gamma , S^1)$ on $\mathscr{H}_{\scriptscriptstyle QC}$ reduces to the representation \eqref{action_M_SC} when restricted to the subgroup $\mathcal{F}\big(M,\widehat{\operatorname{Diff}}_\omega(T^*Q)\big)\subset \operatorname{Diff}(\Gamma )\,\circledS\,\mathcal{F}(\Gamma , S^1)$. Thus, their corresponding momentum maps $J(\Upsilon)=\big(\hbar\operatorname{Im}(\bar \Upsilon{\rm d}\Upsilon),|\Upsilon|^2\big)$ in \eqref{Madelungmomap} and $ \mathcal{D} (\Upsilon)$ in \eqref{Dmomap}, are related by the dual map to the corresponding Lie algebra inclusion $\iota: \mathcal{F}(\Gamma) \hookrightarrow \mathfrak{X}(\Gamma)\,\circledS\, \mathcal{F}(\Gamma)$ given in \eqref{sub_Lie_algebra_inclusion}. Upon denoting $(\sigma,D)=({\cal R}^2\de{\cal S},{\cal R}^2)$ and recalling \eqref{polarD}, this dual map $\iota^*: \mathfrak{X}(\Gamma)^*\times \operatorname{Den}(\Gamma)\rightarrow \operatorname{Den}(\Gamma)$ is computed as
\begin{equation}\label{iota_star}
\iota^*(\sigma_z,\sigma_x, D)= D - \operatorname{div}_z(\mathbb{J}{\mathcal{A}} D - \mathbb{J}\sigma_z)
\,.
\end{equation}
Then, this enables us to write
\[
\iota^*[J(\Upsilon)]= \mathcal{D}(\Upsilon),\quad \text{for all $\Upsilon\in \mathscr{H}_{\scriptscriptstyle QC}$},
\]
which indeed provides an important relation between the momentum map \eqref{Dmomap} for the joint hybrid density and the momentum map $J(\Upsilon)$ in \eqref{Madelungmomap} associated to the hybrid Madelung transform.
\begin{figure}[!h]
\footnotesize\center
\begin{framed}
\begin{xy}
\hspace{.02cm}
\xymatrix{
& & &  & &*+[F-:<3pt>]{
\begin{array}{l}
\vspace{0.1cm}\text{Hybrid density operator \eqref{hybridDenOp}}\\
\vspace{0.1cm} \widehat{ \mathcal{D}}  \in \operatorname{Den}(T^*Q, \operatorname{Her}(\mathscr{H}_{\scriptscriptstyle Q}))\\
\vspace{0.1cm} \text{Hamiltonian $h( \widehat{ \mathcal{D} })$}\\
\end{array}
}
\ar[ddd] |{\begin{array}{c}\text{Map \eqref{Ddef}, dual to the inclusion}\\
\mathcal{F} ( \Gamma ) \hookrightarrow \mathcal{F} (T^*Q, \operatorname{Her}(\mathscr{H}_{\scriptscriptstyle Q}))\end{array}}&\\
& & & & & &\\
& & & & & &\\
&
*+[F-:<3pt>]{
\begin{array}{l}
\vspace{0.1cm}\text{Hybrid wave equation \eqref{hybrid_KvH}}\\
\vspace{0.1cm}\Upsilon \in \mathscr{H}_{\scriptscriptstyle C Q}\\
\vspace{0.1cm}\displaystyle h(\Upsilon)= \int_{T^*Q}\!\big\langle\Upsilon\big|\widehat{\cal L}_{\widehat{H}}\Upsilon\big\rangle\,\Lambda\\
{\rm i}\hbar \partial _t \Upsilon = \widehat{\cal L}_{\widehat{H}}\Upsilon
\end{array}
} \ar[dddrrrr]|{\begin{array}{c}
\text{Momentum map \eqref{Madelungmomap}}\\
\text{for the group}\\
\operatorname{Diff}( \Gamma ) \,\circledS\, \mathcal{F}( \Gamma ,S^1)
\end{array}}
\ar[rrrr]|{\begin{array}{c}\text{Momentum map \eqref{Dmomap}}\\
\text{for the subgroup}\\
\mathcal{F}\big(M, \widehat{\operatorname{Diff}}_\omega(T^*Q)\big)\\
\end{array}}
\ar[ddd]|{\begin{array}{l}\Upsilon= \mathcal{R} e^{{\rm i}  \mathcal{S} /\hbar}\end{array}}
\ar[uuurrrr]& & & & *+[F-:<3pt>]{
\begin{array}{l}
\vspace{0.1cm}\text{Hybrid  density \eqref{Ddef}}\\
\vspace{0.1cm} \mathcal{D}(\Upsilon) \in \operatorname{Den}( \Gamma ) \\
\vspace{0.1cm}\text{Equation  \eqref{CQcont}} 
\end{array}
}\\
& & & & & &\\
& & & & & &\\
&*+[F-:<3pt>]{\begin{array}{l}\text{Hybrid Madelung}\\
\text{equations \eqref{HybHJ}-\eqref{HybHJR} }\\
\text{for $( \mathcal{S} , \mathcal{R} )$}
\end{array}}\ar[rrrr] & & & &
*+[F-:<3pt>]{
\begin{array}{l}
\vspace{0.1cm}\text{Hybrid Madelung variables}\\
\vspace{0.1cm} ( \sigma , D)   \in \big( \mathfrak{X}(\Gamma) \,\circledS\, \mathcal{F}(\Gamma) \big)^*\\ 
\vspace{0.1cm} \text{Hamiltonian $h( \sigma , D)$ in \eqref{collective2}}\\ 
\vspace{0.1cm} \text{Hamiltonian system \eqref{LPEqns1}}  
\end{array}
} \ar[uuu]|{\begin{array}{c}\vspace{0.1cm}\text{Dual to the inclusion } \\
\iota: \mathcal{F}(\Gamma) \hookrightarrow \mathfrak{X}(\Gamma)\,\circledS\, \mathcal{F}(\Gamma)\end{array}} &
}
\end{xy}
\end{framed}
\vspace{-.5cm}\it
\caption{Extended version of Fig. \ref{figure2} including the role of the hybrid density function $\cal D$ and its accompanying geometric structures.}
\label{figure3}
\end{figure}
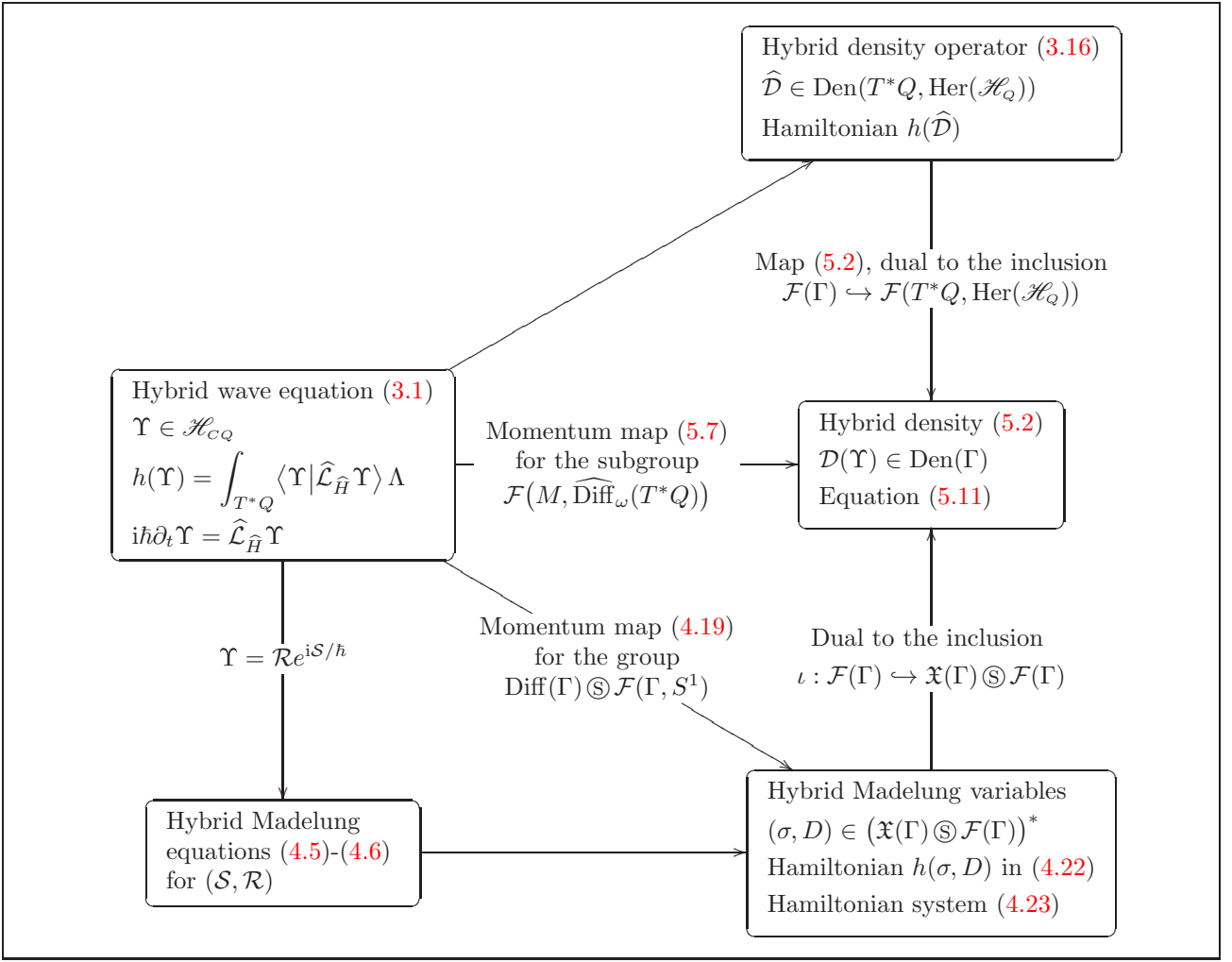
The overall picture is represented in Fig. \ref{figure3}, which extends Fig. \ref{figure2} by presenting the role of the quantum--classical density function $\cal D$ and its relations to the hybrid density operator $\widehat{\cal D}$ and the Madelung variables $(\sigma, D)$. 

\subsection{The quantum--classical continuity equation}\label{continuity_calD}

Although a more geometric picture for the hybrid continuity equation $\partial_t{\cal D}=-\operatorname{div}\bJ$ will be developed in the next section, here we shall simply present the explicit expression of the hybrid current $\bJ(z,x)$  obtained by using the equations \eqref{HybHJ} and \eqref{HybMad2} when taking the time derivative of \eqref{polarD}. This calculation is particularly simplified by noticing that all the terms involving $H_I$ and $L_I$ in \eqref{HybHJ} and \eqref{HybMad2} combine by construction into the right-hand side of equation \eqref{Deq1}. Thus, we can initially drop all the $H_I-$terms from the equations \eqref{HybHJ} and \eqref{HybMad2} (as well as the $L_I-$term in \eqref{HybHJ}) and restore the corresponding term in the ${\cal D}-$equation at a later stage. Upon applying the Leibniz product rule and noticing that
\[
\frac12\bigg\{\R^2,\frac{\Delta_x  \sqrt{\R^2}}{ \sqrt{\R^2}}\bigg\}=\operatorname{div}_x\{\R,{\nabla_x} \R\}
\,,
\]
this process leads to
\begin{equation}\label{CQcont}
\partial_t{\cal D}=-\operatorname{div}{\bf J}=-\operatorname{div}_z J_C-\operatorname{div}_x J_Q
\,,
\end{equation}
with the following classical and quantum component of the hybrid current ${\bf J}=(J_C,J_Q)$:
\begin{align}
J_C:=&\ {\cal D}X_{H_I}\,,\label{J_C}\\
J_Q:=&\ m^{-1}\big(\R^2\nabla_x {\cal S}+\partial_{p_i}(p_i\R^2\nabla_x {\cal S})+\{\R^2\nabla_x  {\cal S},{\cal S}\}
-\hbar^2\{\R,\nabla_x \R\}
\big).\label{J_Q}
\end{align}
We observe that the usual quantum continuity equation is written by simply integrating \eqref{CQcont} over the phase-space coordinates and using $\rho_q(x)=\int_{T^*Q}{\cal D}(z,x)\Lambda=\int_{T^*Q}\R^2(x,z)\Lambda$, thereby obtaining
\[
\frac{\partial \rho_q}{\partial t}= -\frac1m\operatorname{div}_x\int_{T^*Q}(\R^2\nabla_x{\cal S})\Lambda
\,,
\]
whereas the classical density $\rho_c(z)=\int_M{\cal D}(z,x)\mu$ evolves according to
\[
\frac{\partial \rho_c}{\partial t}=\int_M\{H_I,{\cal D}\}\,\mu
\,.
\]

While the geometric origin of the quantum current $J_Q$ will be considered in the next section, here we emphasize that the quantum current $J_Q$ is produced only by the quantum kinetic energy operator $-({\hbar^2}/{2m})\Delta_x$ in the hybrid Hamiltonian \eqref{QHam}, while $H_I$ produces essentially classical dynamics as we discussed in Section \ref{sec:Dmomap}.
Moreover, we point out that it is not known whether $J_Q$ can be divided by $\cal D$ to form a well-defined vector field. This is only  possible if $\cal D$ does not change its sign. As long as $-({\hbar^2}/{2m})\Delta_x$ is retained in \eqref{QHam}, the sign of $\cal D$ is certainly preserved in the absence of quantum--classical coupling (that is $\partial^2_{q^{j\scriptscriptstyle\!} x^k} V_I=0$), as discussed at the end of Section \ref{sec:Dmomap}. However, it is not known whether this happens also in the general case.



\subsection{Hamiltonian structure\label{sec:bracket}}

We have seen that the hybrid Hamiltonian $h(\Upsilon)=\big\langle\Upsilon,\widehat{L}_{\widehat{H}}\Upsilon\big\rangle$, for $\widehat{H}$ given as in \eqref{QHam}, can be written uniquely in terms of $(\sigma, D)$, see \eqref{collective2}.
In order to characterize the continuity equation for $\mathcal{D}$, it is useful to express the hybrid equations in a way that makes $\mathcal{D}$ appear explicitly as an independent variable. To do this, we shall make use of the fact that $\Upsilon\mapsto \mathcal{D}(\Upsilon)$ in \eqref{Dmomap}, or  alternatively, $(\sigma,D) \mapsto \mathcal{D}(\sigma,D)$  in \eqref{iota_star}, are momentum maps and we will apply the following lemma. This produces the explicit Poisson bracket governing the combined dynamics of $\Upsilon$ and ${\cal D}$. In this section, we prefer to express $\Upsilon$ in terms of the variables $(\sigma,D)$.

\medskip
\begin{lemma}\label{lemmaA}
Consider a Poisson manifold $(P,\{\,,\}_P)$, and an equivariant momentum map $\mathbf{J}:P \rightarrow \mathfrak{g}^*$  with respect to a left canonical action of the Lie group $G$ on $P$. Then the map
\[
P \rightarrow \mathfrak{g}^* \times P,\qquad p \mapsto (\mathbf{J}(p),p)
\]
is a Poisson map with respect to the Poisson bracket $\{\,,\}_P$ on $P$ and the Poisson bracket
\begin{equation}\label{Poisson}
\{f,g\}= \{f,g\}_+ + \{f,g\}_P - \left\langle \frac{\partial g}{\partial p}, \Big(\frac{\partial f}{\partial \sigma}\Big)_P \right\rangle + \left\langle \frac{\partial f}{\partial p}, \Big(\frac{\partial g}{\partial \sigma}\Big)_P \right\rangle
\end{equation}
on $\mathfrak{g}^* \times P$.
In particular, given a Hamiltonian $H: P \rightarrow\mathbb{R}$, if $p(t)\in P$ is a solution of Hamilton's equations for $H$, then $(\nu(t),p(t))=(\mathbf{J}(p(t)), p(t))\in \mathfrak{g}^*\times P$ is a solution of Hamilton's equation for $h:\mathfrak{g}^*\times P\to\Bbb{R}$ with respect to the Poisson bracket \eqref{Poisson}, where $h$ is a function such that $h(\mathbf{J}(p),p)= H(p)$, for all $p\in P$. 
\end{lemma}
\paragraph{Proof.} This follows from a result of \cite{KrMa2987} (see Prop. 2.2 therein) stating that the map 
\begin{equation}\label{KM}
(\nu, p ) \in \mathfrak{g}^*\times P \mapsto(\nu +\mathbf{J}(p), p) \in \mathfrak{g}^*\times P
\end{equation}
is a Poisson diffeomorphism sending the Poisson bracket $\{\,,\}_++ \{\,,\}_P$ to the Poisson bracket \eqref{Poisson}.
Then, we note that $(P, \{\,,\}_P)$ is a Poisson submanifold of $(\mathfrak{g}^*\times P,\{\,,\}_++ \{\,,\}_P)$ with respect to the inclusion $p\mapsto (0,p)$, as a direct computation shows. By composing this inclusion with the Poisson map \eqref{KM}, the result follows. $\quad\quad\blacksquare$

\medskip
\noindent
This result can be applied to $P=\mathscr{H}_{\scriptscriptstyle QC}\ni\Upsilon$ endowed with the symplectic Poisson bracket. In this case the Lie group $G=\mathcal{F}\big(M, \widehat{\operatorname{Diff}}_\omega(T^*Q)\big)$ acts canonically on the left as in \eqref{action_M_SC} with associated equivariant momentum map $\Upsilon \mapsto \mathcal{D}(\Upsilon)$ given in \eqref{Dmomap}. Alternatively, we have the following result.

\begin{theorem}[Hamiltonian structure]
With the definitions \eqref{J_C}-\eqref{J_Q} and $(\sigma,D):=({\cal R}^2\de{\cal S},{\cal R}^2)$, the system comprised by equations \eqref{LPEqns1} and \eqref{CQcont} is Lie-Poisson for the semidi-rect-product group $G\,\circledS\,S$, with $G=\mathcal{F}\big(M, \widehat{\operatorname{Diff}}_\omega(T^*Q)\big)$, $S=\operatorname{Diff}(\Gamma)\,\circledS\,\mathcal{F}(\Gamma,S^1)$, and the inclusion $G\subset S$ given in Lemma \ref{subgrouplemma}. In particular, if $\iota:\mathfrak{g}\hookrightarrow\mathfrak{s}$ is the corresponding Lie algebra inclusion and
\beq\label{adstar}
\operatorname{ad}^*_{\left(v, f\right)}(\sigma,D)=\big( \pounds_{v}\sigma + D\nabla f, \,\operatorname{div}_x(Dv_x)\big)
\eeq
is the infinitesimal coadjoint action for $\mathfrak{s}$ with respect to the standard $L^2$ duality pairing $\langle\cdot,\cdot\rangle:\mathfrak{s}^*\times\mathfrak{s}\to\Bbb{R}$,
then  equations \eqref{LPEqns1} and \eqref{CQcont} are Lie-Poisson with respect to the bracket
\begin{multline}\label{LPB2}
\{f,h\}({\cal D},\sigma,D)=  \int _\Gamma\!\bigg[\,{\cal D} \left\{\frac{\delta f}{\delta {\cal D}} ,  \frac{\delta h}{\delta {\cal D}}\right\}
- \sigma_j \left( \frac{\delta f}{\delta \sigma } \cdot \nabla  \frac{\delta h}{\delta \sigma_j}
- \frac{\delta h}{\delta \sigma } \cdot \nabla  \frac{\delta f}{\delta \sigma_j} \right)
\\
+ D \left( \frac{\delta f}{\delta \sigma } \cdot \nabla  \frac{\delta h}{\delta D}
- \frac{\delta h}{\delta \sigma } \cdot \nabla  \frac{\delta f}{\delta D} \right)\!\bigg]\mu_\Gamma\\
+
\left\langle\bigg(\frac{\delta h}{\delta \sigma},\frac{\delta h}{\delta D}\bigg),\operatorname{ad}^*_{\iota\left({\delta f}/{\delta {\cal D}}\right)}(\sigma,D)\right\rangle
-
\left\langle\bigg(\frac{\delta f}{\delta \sigma},\frac{\delta f}{\delta D}\bigg),\operatorname{ad}^*_{\iota\left({\delta g}/{\delta  {\cal D}}\right)}(\sigma,D)\right\rangle
\,
\end{multline}
and the Hamiltonian 
\beq\label{giacomo}
h(\sigma, D,{\cal D})= \int_\Gamma\bigg[\frac{1}{2m}\frac{|\sigma_x|^2}{D} + \frac{\hbar^2}{8m}\frac{|\nabla_x D|^2}{D} + H_I {\cal D} \bigg]\mu_\Gamma\,.
\eeq
\end{theorem}
\paragraph{Proof.}
In Lemma \ref{lemmaA}, we choose $P=\big(\mathfrak{X}(\Gamma)\,\circledS\,\mathcal{F}(\Gamma)\big)^*\ni (\sigma,D)$ endowed with the Lie-Poisson bracket \eqref{LPB}. Then, $G=\mathcal{F}\big(M, \widehat{\operatorname{Diff}}_\omega(T^*Q)\big)$ acts canonically from the left on $P$ by the coadjoint action of $\operatorname{Diff}(\Gamma)\,\circledS\,\mathcal{F}(\Gamma,S^1)$, suitably restricted to the subgroup $G$. See the discussion at the end of Section  \ref{sec:Dmomap}. 
Notice that the Poisson manifold $P$ is the dual space $\mathfrak{s}^*$ of the semidirect-product Lie algebra $\mathfrak{s}=\mathfrak{X}(\Gamma)\,\circledS\,\mathcal{F}(\Gamma)$ of the automorphism group ${S}=\operatorname{Aut}(\Gamma\times S^1)=\operatorname{Diff}(\Gamma)\,\circledS\,\mathcal{F}(\Gamma,S^1)$ of the trivial circle bundle $\Gamma\times S^1\to\Gamma$. Then, since $G=\mathcal{F}\big(M, \operatorname{Aut}_{\cal A}(T^*Q\times S^1)\big)$ 
is a subgroup of $S$, the corresponding group inclusion generates a semidirect-product structure $G\,\circledS\,S$, so that Lemma \ref{lemmaA} leads to a Lie-Poisson bracket on the dual Lie algebra $(\mathfrak{g}\,\circledS\,\mathfrak{s})^*$. In this context, the momentum map associated to subgroup action of $G$ on ${S}$ is given by \eqref{iota_star}, that is the dual of the Lie algebra inclusion $\iota:\mathfrak{g}\hookrightarrow\mathfrak{s}$. In this case, the infinitesimal generator associated to the Lie algebra element $\xi \in \mathfrak{g}=\mathcal{F}(\Gamma)$ acting on $(\sigma,D)\in P=\mathfrak{s}^*$ reads $- \operatorname{ad}^*_{\iota(\xi)}(\sigma,D)$, where $\operatorname{ad}^*$ is the infinitesimal coadjoint action \eqref{adstar}.
Using this, the Lie-Poisson bracket \eqref{Poisson} on $\big(\mathcal{F}(\Gamma)\,\circledS\big(\mathfrak{X}(\Gamma)\,\circledS\,\mathcal{F}(\Gamma)\big)\big)^*$ gives \eqref{LPB2}.

The hybrid Hamiltonian $h(\Upsilon)$ in \eqref{hybHam}, with $\widehat{H}$ given as in \eqref{QHam}, can be written in terms of $(\sigma, D)$ and ${\cal D}=\iota^*(\sigma,D)$ as \eqref{giacomo}. 
Then, the bracket \eqref{LPB2} yields the  Lie-Poisson equations
\begin{align*}
(\partial_t\sigma, \partial_tD)&= -\operatorname{ad}^*_{\left(\frac{\delta h}{\delta \sigma}, \frac{\delta h}{\delta D}\right)}(\sigma,D) - \operatorname{ad}^*_{\iota\left( \frac{\delta h}{\delta {\cal D}}\right)}(\sigma,D)
\\
\partial_t{\cal D}&= -\Big\{ {\cal D},\frac{\delta h}{\delta {\cal D}}\Big\}- \iota ^*\Big(\operatorname{ad}^*_{\left(\frac{\delta h}{\delta \sigma}, \frac{\delta h}{\delta D}\right)}\Big)(\sigma,D)\,,
\end{align*}
The first line above recovers the equations \eqref{LPEqns1}. This follows from evaluating ${\delta h}/{\delta\sigma_z}=0$ and ${\delta h}/{\delta {\cal D}}=H_I$ and recalling the expression  $\iota(\xi)=((X_\xi,0), -\xi_{\mathcal{A}})$ of the Lie algebra inclusion (see the end of Section \ref{sec:Dmomap}).
On the other hand, the quantum--classical continuity equation \eqref{CQcont} emerges from the ${\cal D}-$equation above by recognizing that the two terms on the right-hand side
  identify exactly the contributions of the classical and quantum currents as
\begin{equation}\label{JC_JQ}
\left\{{\cal D}, H_I\right\}= \operatorname{div}_z J_C\qquad\text{and}\qquad \iota ^*\Big(\operatorname{ad}^*_{\left(\frac{\delta h}{\delta \sigma}, \frac{\delta h}{\delta D}\right)}\Big)(\sigma,D)= \operatorname{div}_x J_Q\,.
\end{equation}
The proof is completed after verifying the second equality as in Appendix \ref{A}. $\qquad\qquad\blacksquare$

\bigskip
\noindent
Notice  that if the quantum kinetic energy operator $-({\hbar^2}/{2m})\Delta_x$ is absent in the hybrid Hamiltonian \eqref{QHam}, then the Hamiltonian $h(\Upsilon)$ in \eqref{hybHam} \emph{collectivizes} (in the sense of Guillemin and Sternberg \cite{Sternberg2}) with respect to the momentum map \eqref{Dmomap}. Indeed, the previous expression \eqref{giacomo} reduces to
\[
h(\Upsilon)= \int_{T^*Q}\operatorname{Tr}(\widehat{H}(z)\widehat{\mathcal{D}}(z))\Lambda=\int_\Gamma  H_I(z,x)\mathcal{D}(z,x)\mu_\Gamma .
\]
In this case, the Lie-Poisson equations decouple thereby recovering  the previous result  \eqref{Deq1}.

\subsection{A class of Hamiltonians preserving positivity\label{sec:Hamclass}}
In this section we identify an infinite family of hybrid systems for which both the quantum density matrix and the classical Liouville density are positive in time. Indeed, while the quantum density matrix \eqref{QuantDensMat} is always positive-definite by construction, the sign of the classical Liouville density \eqref{rhoc} requires further study.
In this section, we shall consider hybrid Hamiltonians of the form
\beq
\label{classHam}
\widehat{H}(z)=H(z,\widehat{\alpha})\,,
\eeq
where $\widehat{\alpha}$ is a purely quantum observable, i.e. it is an Hermitian operator on $\mathscr{H}_{\scriptsize Q}$.
Here, we assume that the dependence of $H$ on $\widehat{\alpha}$ is analytic. As we shall see, any hybrid wave equation \eqref{hybrid_KvH} associated to the type of Hamiltonian \eqref{classHam} leads to the positivity of both quantum and classical densities. In what follows, it is convenient to use Dirac's notation so that, upon recalling the isomorphism \eqref{isom1}, a hybrid state vector is denoted by
\[
|\Upsilon(z)\rangle:=\Upsilon(z) \in\mathscr{H}_{\scriptsize C}\otimes\mathscr{H}_{\scriptsize Q}
\,.
\]
By a slight generalization of \eqref{classHam}, the following statement replaces $\widehat\alpha$ by any set of mutually commuting quantum observables.

\begin{proposition}
Consider  a hybrid Hamiltonian $\widehat{H}(z)$ depending on a sequence $\{\widehat{\alpha}_i\}_{i=1,\dots,N}$ of mutually commuting quantum observables, that is
\beq
\widehat{H}(z)=H(z,\{\widehat{\alpha}_i\})\,,
\label{classHam2}
\eeq
where $[\widehat{\alpha}_i,\widehat{\alpha}_j]=0$ for all $i,j=1,\dots, N$. Also, consider a solution  $|\Upsilon(z,t)\rangle $ of the associated quantum--classical wave equation \eqref{hybrid_KvH} and define the projection $\Upsilon(z,\alpha,t)=\langle\alpha|\Upsilon(z,t)\rangle$. Then, the sign of the joint distribution
\beq
\widetilde{\cal D}(z,\alpha,t):= |\Upsilon(z,\alpha,t)|^2-  \partial_{p_i} \big(p_i|\Upsilon(z,\alpha,t)|^2\big) + {\rm i}\hbar \{\Upsilon(z,\alpha,t), \bar\Upsilon(z,\alpha,t)\}
\label{Liuda}
\eeq
is preserved in time.
\end{proposition}
\paragraph{Proof.} Since mutually commuting Hermitian operators share the same eigenvalue problem, here we loose no generality by restricting to the case \eqref{classHam} of only one quantum observable $\widehat{\alpha}\in\operatorname{Her}(\mathscr{H}_{\scriptsize Q})$.
Upon denoting $\Lambda=-{\rm i}\hbar\nabla_z$, the quantum--classical wave equation \eqref{hybrid_KvH} reads
\[
{\rm i}\hbar\partial_t|\Upsilon(z)\rangle=X_H(z,\widehat{\alpha})\cdot\Lambda|\Upsilon(z)\rangle-{L}(z,\widehat{\alpha})|\Upsilon(z)\rangle
\,,
\]
where $L(z, \widehat{\alpha})= \mathcal{A}\cdot X_H(z, \widehat{\alpha})- H(z, \widehat{\alpha})=\sum_n L_n(z)\widehat{\alpha}^n$.
Now,  consider the spectrum of $\widehat{\alpha}$, that is $\widehat{\alpha}|\alpha\rangle=\alpha|\alpha\rangle$ and write
\begin{equation}\label{hybrid_alpha}
{\rm i}\hbar\partial_t\langle\alpha|\Upsilon(z)\rangle=\int\!\Big( \langle\alpha|X_H(z,\widehat{\alpha})|\alpha'\rangle\cdot\Lambda\langle\alpha'|\Upsilon(z)\rangle - \langle\alpha|{L}(z,\widehat{\alpha})|\alpha'\rangle\langle\alpha'|\Upsilon(z)\rangle\Big)\de\alpha'
\,,
\end{equation}
where $\langle\alpha|:=|\alpha\rangle^\dagger$
for all $\alpha$.
The term $\langle\alpha|{L}(z,\widehat{\alpha})|\alpha'\rangle$ can be rewritten according to
\begin{align*}
\langle\alpha|{L}(z,\widehat{\alpha})|\alpha'\rangle=&\ 
\sum_n{L}_n(z)\langle\alpha|\widehat{\alpha}^n|\alpha'\rangle
\\
=&\ 
\sum_n{L}_n(z)({\alpha}')^n\langle\alpha|\alpha'\rangle
\\
=&\ 
{L}(z,{\alpha}')\delta(\alpha-\alpha')
\\
=&\ 
\big(\mathcal{A}(z) \cdot X_H(z,{\alpha}') - H(z,\alpha')\big)\delta(\alpha-\alpha')
\end{align*}
and by proceeding analogously one also has
$
\langle\alpha|X_H(z,\widehat{\alpha})|\alpha'\rangle=X_H(z,{\alpha}')\delta(\alpha-\alpha')
$.
Then, upon writing $\Upsilon(z,\alpha):=\langle\alpha|\Upsilon(z)\rangle$, equation \eqref{hybrid_alpha} becomes
\begin{align*}
{\rm i}\hbar\partial_t\Upsilon(z,\alpha)=&\,X_H(z,{\alpha})\cdot\Lambda\Upsilon(z,\alpha) 
-\big(\mathcal{A}\cdot X_H(z,{\alpha})-H(z,\alpha)\big)\Upsilon(z,\alpha)
\\
=&\,\widehat{{\cal L}}_{H(z,{\alpha})}\Upsilon(z,\alpha)
\,.
\end{align*}
At this point, we construct the joint quantum--classical density \eqref{Liuda}
for the classical position $z$ in phase-space and the quantum degree of freedom $\alpha$. Upon following the same arguments as in Section \ref{sec:Dmomap}, one shows that the joint density $\widetilde{\cal D}(z,\alpha)$ is a momentum map $L^2(T^*Q\times \sigma(\widehat{\alpha}))\to\operatorname{Den}(T^*Q\times \sigma(\widehat{\alpha}))$, where $\sigma(\widehat{\alpha})$ denotes the spectrum of $\widehat{\alpha}$. Consequently, the hybrid density $\widetilde{\cal D}(z,\alpha)$ satisfies the Liouville equation
\[
\partial_t\widetilde{\cal D}(z,\alpha)=\{H,\widetilde{\cal D}\}(z,\alpha)=-\pounds_{X_H}\widetilde{\cal D}
\,,
\]
which indeed preserves the sign of $\widetilde{\cal D}(z,\alpha)$.
Thus, if at the initial time
\begin{equation}\label{initial_condition}
\widetilde{\cal D}(t=0,z,\alpha)\geq 0
\,, \qquad\ 
\forall\;(z,\alpha)\in T^*Q\times \sigma(\widehat{\alpha}),
\end{equation}
then $\widetilde{\cal D}(t,z,\alpha)\geq 0$ for all times and for all $(z,\alpha)\in T^*Q\times \sigma(\widehat{\alpha})$. $\qquad\qquad\blacksquare$

\medskip\noindent 
For example, $\widetilde{\cal D}(t=0,z,\alpha)$ is positive  whenever the hybrid density operator $\widehat{\cal D}(z)$ is positive-definite at the initial time, that is
\[
\langle \psi |\widehat{\cal D}(t=0,z)|\psi\rangle \geq 0
\,,\qquad
\forall\;|\psi\rangle \in\mathscr{H}_{\scriptscriptstyle Q}\,.
\]
Since the classical density in \eqref{luigi} can be written as $\rho_c(t,z)=\int\mathcal{D}(t,z,x)\mu={\int \widetilde{\mathcal{D}}(t,z,\alpha){\rm d}\alpha\geq 0}$,  then we obtain the following result:
\begin{corollary}
Assume that the hybrid density operator $\widehat{\cal D}(z)$ is positive at the initial time, then the density $\rho_c$ is also positive at initial time and its sign is preserved by the quantum--classical wave equation \eqref{hybrid_KvH} with Hamiltonian of the type \eqref{classHam2}.
\end{corollary}

In the simplest case, we can consider the position operator $\widehat{\bx}$ on  $M=\Bbb{R}^n$ such that  $[\widehat{x}_i,\widehat{x}_j]=0$; then, one recovers the results in Section \ref{sec:Dmomap} for the joint probability density $\mathcal{D}(z,\bx)$. Analogously, one can consider the momentum operator $\widehat{\boldsymbol{p}}=-{\rm i}\hbar\nabla_x$ so that $[\widehat{p}_i,\widehat{p}_j]=0$ and construct a Hamiltonian of the type
\[
\widehat{H}(z)=\widehat{H}(z,\widehat{\boldsymbol{p}})
\,.
\]
In this case the eigenvectors are $|\boldsymbol{k}\rangle= (2\pi\hbar)^{-n/2\,}e^{{\rm i}\boldsymbol{k} \cdot \bx/\hbar}$ and $\Upsilon(z,\boldsymbol{k})=\langle \boldsymbol{k}|\Upsilon(z)\rangle$  are the  Fourier transforms
\[
\Upsilon(z,\boldsymbol{k})=\frac{1}{\big(\sqrt{2\pi\hbar}\,\big)^n}\int_M\Upsilon(z,\bx)e^{-{\rm i}\boldsymbol{k}\cdot \bx/\hbar}\mu.
\]
Another case of possible interest is that  of a finite-dimensional quantum Hilbert space $\mathscr{H}_{\scriptscriptstyle Q}$, for which one repeats the same steps and eventually is left with
\[
{\rm i}\hbar\partial_t\Upsilon_n=
{\cal L}_{H({\alpha_n})}\Upsilon_n
\]
so that the density
\[
\widetilde{\cal D}_n(z):=|\Upsilon_n(z)|^2-  \partial_{p_i} \big(p_i|\Upsilon_n(z)|^2\big)+{\rm i}\hbar\{\Upsilon_n,\bar\Upsilon_n\}(z)
\]
satisfies the Liouville equation $\partial_t\widetilde{\cal D}_n(z)=\{H(z,\alpha_n),\tilde{\cal D}_n(z)\}$ and thus the same conclusion as in the continuum case holds for the classical density $\rho_c=\sum_n\widetilde{\cal D}_n$. In the case $\mathscr{H}_{\scriptscriptstyle Q}=\Bbb{C}^2$ of two-level quantum subsystems, a proof of this result already appeared in \cite{BoGBTr}.

\section{Conclusions}

Despite the absence of classical particle trajectories in quantum--classical dynamics, this paper has addressed the problem of identifying a Hamiltonian flow governing the motion of the classical subsystem within the entire hybrid system. In more generality, hybrid  Bohmian trajectories were identified by applying the Madelung-Bohm picture to the quantum--classical  wavefunction $\Upsilon(q,p,x)$. In addition, the continuity equation \eqref{CQcont} for the quantum--classical density \eqref{Ddef} was presented explicitly, along with the hybrid current distribution \eqref{J_C}-\eqref{J_Q} extending the  probability current from standard quantum mechanics. 

The results in this paper shed a new light on the 40-year old problem of quantum--classical coupling. Indeed, while several general ideas about phenomenological aspects have emerged over a century of continuing efforts, a mathematical foundation of quantum measurement is still absent. A theory of quantum--classical coupling represents a relevant step forward as a prelude to a measurement theory.  For example, hybrid Bohmian trajectories may lead to a new understanding of the measurement process without the need of invoking the wavefunction collapse postulate, which is indeed avoided in the pilot-wave interpretation of standard quantum mechanics \cite{Bohm}.
Alternatively,   hybrid Bohmian trajectories may also be used to design new reduced models for  nonadiabatic molecular dynamics (see \cite{FoHoTr} for a geometric hydrodynamic treatment thereof), of paramount importance in chemical physics. In this context, the difficulties of a full quantum treatment lead to the necessity of modeling nuclei as classical particles while retaining the full quantum treatment of electron dynamics. Such models are typically formulated by taking semiclassical limits of a full quantum treatment and in most cases this process suffers from not capturing the \emph{quantum backreaction} beyond mean-field effects. As the quantum backreaction is intrinsically built in the approach formulated in this paper, hybrid Bohmian trajectories may serve as a point of departure for formulating closure models overcoming the issues present in conventional molecular dynamics simulations. We intend to develop this particular direction in the near future.


The present hybrid theory is formulated by starting from the Koopman-van Hove equation for two classical particles and then applying a partial quantization procedure leading to the  quantum--classical wavefunction $\Upsilon(q,p,x)$, where $(q,p)$ are classical phase-space coordinates while $x$ is the coordinate on the quantum configuration space. This wavefunction undergoes a unitary evolution generated by a {hybrid Liouvillian operator} associated to the quantum--classical Hamiltonian. The long-sought equivariance properties  of hybrid Liouvillians under both quantum and classical transformations were studied in Section \ref{sec:algebra}, which also presented a remarkable relation relating commutators and Poisson brackets. Moreover, Section \ref{sec:hybden} formulated a hybrid density operator extending the quantum density matrix to the quantum--classical setting; while  the  density matrix of the quantum subsystem is always positive-definite by construction, the hybrid quantum--classical density is generally allowed to be unsigned and this point was developed further in the second part of the paper.

In Section \ref{sec:HybMad}, we applied the symplectic geometry of the Madelung transform to hybrid wavefunctions and obtained fluid-like  Lagrangian paths providing a hybrid quantum--classical extension of the celebrated Bohmian trajectories in quantum mechanics. In the presence of quantum--classical coupling, the symplectic form on the classical phase-space is not preserved by the hybrid flow and explicit equations of motion were presented for the Poincar\'e integral, which is no longer a dynamical invariant. Nevertheless, the classical phase-space components of the hybrid Bohmian trajectories identify a Hamiltonian flow parameterized by the quantum coordinate. This flow is associated to the motion of the classical subsystem and it was indeed shown to preserve the classical symplectic form. In addition, the Hamiltonian  structure of the hybrid Madelung equations was also characterized explicitly in terms of reduction by symmetry in Section \ref{Ham_var}.

In the last part of the paper, the joint quantum--classical density is considered in terms of its underlying momentum map structure. A hybrid continuity equation was presented in Section \ref{continuity_calD}, thereby identifying hybrid quantum--classical current mimicking the quantum probability current. The hybrid continuity equation \eqref{CQcont} and its current distribution \eqref{J_C}-\eqref{J_Q} were also shown to emerge from a Lie-Poisson Hamiltonian structure, which sheds more light on the geometry underling the hybrid  density evolution. While the latter does not generally preserve the sign of the distribution, the paper concludes by characterizing an infinite family of hybrid systems preserving the sign of the classical probability density.

\addtocontents{toc}{\protect\setcounter{tocdepth}{0}}
\appendix

\section{Quantum component of the hybrid current}\label{A}

Here, we prove the second equality in \eqref{JC_JQ}. Using the expression of $\iota^*$ in \eqref{iota_star}, we get
\begin{align*}
\iota ^*\operatorname{ad}^*_{\left(\frac{\delta h}{\delta \sigma}, \frac{\delta h}{\delta D}\right)}(\sigma,D)&= \iota ^*\left( \pounds_{\frac{\delta h}{\delta \sigma}}\sigma + D\nabla \frac{\delta h}{\delta D}, \operatorname{div}_x\big(D\frac{\delta h}{\delta \sigma_x}\big)\right)\\
&= \operatorname{div}_x\left(D\frac{\delta h}{\delta \sigma_x}\right) - \operatorname{div}_z\left(\mathbb{J}\mathcal{A} \operatorname{div}_x\Big(D\frac{\delta h}{\delta \sigma_x}\Big)- \mathbb{J}\Big( \pounds_{\frac{\delta h}{\delta \sigma}}\sigma + D\nabla \frac{\delta h}{\delta D}\Big)_z\right)\,.
\end{align*}
Since $D\,{\delta h}/{\delta \sigma_x}= m^{-1} \R^2{\rm d}_x {\cal S}$, the first term is $m^{-1}\operatorname{div}_x \big(\R^2{\rm d}_x {\cal S})$. The second term is
\[
-\operatorname{div}_z\left(\mathbb{J}\mathcal{A} \operatorname{div}_x\Big(D\frac{\delta h}{\delta \sigma_x}\Big)\right)
=
-\operatorname{div}_x\left(\operatorname{div}_z\Big(\mathbb{J}\mathcal{A} D\frac{\delta h}{\delta \sigma_x}\Big)\right)= m^{-1}\operatorname{div}_x\big(\partial_{p_i}(p_i\R^2\nabla_x S)\big).
\]
Then, we apply $\operatorname{div}_z\Bbb{J}$ to the $z$-component of $ \pounds_{{\delta h}/{\delta \sigma}}\sigma $, with ${\delta h}/{\delta\sigma_z}=0$. We find
\begin{align*}
\operatorname{div}_z\left[\Bbb{J}\left(
\operatorname{div}\Big(\frac{\delta h}{\delta \sigma}\otimes\sigma_z\Big)
+
\Big(\nabla_z\frac{\delta h}{\delta \sigma}\Big)\cdot\sigma
\right)\right]
=&\ 
\operatorname{div}_z\left[\left(
\operatorname{div}_x\Big(\frac{\delta h}{\delta \sigma_x}\otimes\Bbb{J}\sigma_z\Big)
+
\Bbb{J}\left(\nabla_z\frac{\delta h}{\delta \sigma_x}\right)\cdot\sigma_x
\right)\right]
\\
=&\ 
\operatorname{div}_x\left[\operatorname{div}_z
\Big(\Bbb{J}\sigma_z\otimes\frac{\delta h}{\delta \sigma_x}\Big)\right]
+
\operatorname{div}_z\left[\Bbb{J}
\Big(\nabla_z\frac{\delta h}{\delta \sigma_x}\Big)\cdot\sigma_x
\right]\\
=&\operatorname{div}_x\left[\operatorname{div}_z
\Big(\Bbb{J}\nabla_z{\cal S} \otimes\frac{\R^2}{m}\nabla_x{\cal S}\Big)\right] + \left\{\sigma_x,\frac{\delta h}{\delta \sigma_x}\right\}\\
=&\ m^{-1}\operatorname{div}_x \{\R^2\nabla_x {\cal S}, {\cal S}\}+\left\{\sigma_x,\frac{\delta h}{\delta \sigma_x}\right\}\,.
\end{align*}
Finally, we compute
\[
\operatorname{div}_z\Bbb{J}\left(D\nabla_z \frac{\delta h}{\delta D}\right)=\left\{D,\frac{\delta h}{\delta D}\right\}\,.
\]
The result follows by noting that
\begin{align*}
 \left\{\sigma_x,\frac{\delta h}{\delta \sigma_x}\right\}
+
\left\{D,\frac{\delta h}{\delta D}\right\}=&\ 
\left\{\sigma_x,\frac{\sigma_x}{mD}\right\}-\frac1{2m}\left\{D,\Big(\frac{\sigma_x}D\Big)^2\right\}+
\frac12\left\{D,\frac{\Delta_x  \sqrt{D}}{ \sqrt{D}}\right\}
\\
=&\ 
\operatorname{div}_x\{\sqrt{D},{\nabla_x}\sqrt{D}\}
\end{align*}
and using the expression \eqref{J_Q} for $J_Q$.

\section{Proof of the equivariance lemma \ref{lemmaequiv}\label{app:equiv}}
In this Appendix, we present a proof the equivariance property \eqref{ignazio} of the hybrid Liouvillian under strict contact transformations. Upon using the notation $\varphi(z)$ defined on the right-hand side of \eqref{Aut-relations}, we write
\begin{align*}
&\left(U_{(\eta, e^{{\rm i}\theta})}^\dagger \widehat{\mathcal{L}}_{\widehat{A}}U_{(\eta, e^{{\rm i}\theta})}\Upsilon\right)(z)
\\
&= \left(\widehat{\mathcal{L}}_{\widehat{A}}U_{(\eta, e^{{\rm i}\theta})}\Upsilon \right)(\eta(z))\exp \big( {\rm i}\hbar^{-1} \varphi(z)\big)\\
&= \Big({\rm i}\hbar \big\{\widehat{A}, \Upsilon\circ\eta^{-1} \exp \big(- {\rm i}\hbar^{-1}\varphi\circ\eta^{-1}\big)\big\}(\eta(z)) \\
&\qquad - \big(\mathcal{A}(\eta(z))\cdot X_{\widehat{A}}(\eta(z))- \widehat{A}(\eta(z))\big)\Upsilon(z)\exp\big(-{\rm i} \hbar^{-1} \varphi(z)\big)\Big)\exp \big( {\rm i}\hbar^{-1} \varphi(z)\big)\\
&= {\rm i}\hbar \{\widehat{A}, \Upsilon\circ\eta^{-1}\}(\eta(z)) + {\rm i}\hbar \{ \widehat{A}, \exp\big(-{\rm i} \hbar^{-1} \varphi\circ\eta^{-1}\big)\}(\eta(z))\Upsilon(z) \exp \big( {\rm i}\hbar^{-1} \varphi(z)\big)\\
&\qquad - \big(\mathcal{A}\cdot X_{\widehat{A}\circ\eta(z)}- \widehat{A}\circ\eta)\big)(z)\Upsilon(z) + {\rm d}\varphi (z) \cdot X_{\widehat{A}\circ\eta(z)}\Upsilon(z)\\
&= \Big(  \widehat{\mathcal{L}}_{\widehat{A}\circ\eta}\Upsilon\Big)(z) + {\rm i}\hbar \{\widehat{A}\circ \eta, \exp\big(-{\rm i} \hbar^{-1} \varphi\big) \}(z)\Upsilon(z)  \exp\big({\rm i} \hbar^{-1} \varphi(z)\big) + {\rm d}\varphi (z) \cdot X_{\widehat{A}\circ\eta(z)}\Upsilon(z)\\
&= \Big(  \widehat{\mathcal{L}}_{\widehat{A}\circ\eta}\Upsilon\Big)(z) - {\rm i}\hbar \,{\rm d} \big(\exp\big(-{\rm i} \hbar^{-1} \varphi\big)\big)(z)\cdot X_{\widehat{A}\circ\eta}(z) \Upsilon(z)  \exp\big({\rm i} \hbar^{-1} \varphi(z)\big) + {\rm d}\varphi (z) \cdot X_{\widehat{A}\circ\eta(z)}\Upsilon(z)\\
&= \Big(  \widehat{\mathcal{L}}_{\widehat{A}\circ\eta}\Upsilon\Big)(z).
\end{align*}
In the third equality we used $\eta^*\mathcal{A}+\de\varphi=\mathcal{A}$ and we emphasize that the symplectic potential $\mathcal{A}$ should not be confused with the hybrid observable $\widehat{A}\in \mathcal{F}(T^*Q,\operatorname{Her}(\mathscr{H}_{\scriptscriptstyle Q}))$.

\smallskip
\paragraph{Acknowledgements.} This material is partially based upon work supported by the NSF Grant No. DMS-1440140 while CT was in residence at the MSRI, during the Fall 2018 semester. In addition, CT acknowledges support from the Alexander von Humboldt Foundation (Humboldt Research Fellowship for Experienced Researchers) as well as from the German Federal Ministry for Education  and  Research. FGB is partially supported by the ANR project GEOMFLUID, ANR-14-CE23-0002-01.

\end{document}